\documentclass[number,sort&compress]{elsarticle}

\usepackage{graphicx} 
\usepackage{dcolumn}  
\usepackage{subfigure} 
\usepackage{amsmath}
\usepackage{varioref}
\usepackage{multirow}
\graphicspath{{ps}}
\usepackage{grffile} 
\usepackage{color}
\usepackage{listings}
\definecolor{ourEm}{rgb}{0,0,0} 

\begin{document}



\title{ \quad\\[1.0cm] A Hierarchical NeuroBayes-based Algorithm for Full Reconstruction of $B$ Mesons at B Factories}



\address[EKP]{Institut f\"ur Experimentelle Kernphysik, Karlsruher Institut f\"ur Technologie, Campus S\"ud, Postfach 69 80, 76128 Karlsruhe}
\address[Warwick]{now at Department of Physics, University of Warwick, Coventry, CV4 7AL}
 \author[EKP]{M.~Feindt}
 \author[EKP]{F.~Keller}
 \author[EKP,Warwick]{M.~Kreps}
 \author[EKP]{T.~Kuhr}
 \author[EKP]{S.~Neubauer}
 \author[EKP]{D.~Zander}
 \author[EKP]{A.~Zupanc}

\begin{abstract}
We describe a new $B$-meson full reconstruction algorithm designed for the Belle experiment at the 
B-factory KEKB, an asymmetric $e^+e^-$ collider that collected a data sample of $771.6 \times 10^{6}$
$B\bar{B}$ pairs during its running time. 
To maximize the number of reconstructed $B$ decay 
channels, it utilizes a hierarchical reconstruction procedure and probabilistic calculus instead
of classical selection cuts.
The multivariate analysis package NeuroBayes was used extensively to hold the 
balance between highest possible efficiency, robustness and acceptable consumption of CPU time.

In total, 1104 exclusive decay channels were reconstructed, employing 71 neural
networks altogether. Overall, we correctly reconstruct one $B^{\pm}$ or $B^0$ candidate in $0.28\%$ or $0.18\%$ of the $B\bar{B}$ events, respectively. Compared to the cut-based classical reconstruction algorithm used at the Belle experiment, 
this is an improvement in efficiency by roughly a factor of $2$, depending on the analysis considered.

The new framework also features the ability to choose the desired purity or efficiency of the fully reconstructed
sample freely. If the same purity as for the classical full reconstruction code is desired ($\sim 25\%$), the efficiency
is still larger by nearly a factor of $2$. If, on the other hand, the efficiency is chosen at a similar level as the 
classical full reconstruction, the purity rises from $\sim 25\%$ to nearly $90\%$.

\end{abstract}


\maketitle


{\renewcommand{\thefootnote}{\fnsymbol{footnote}}}
\setcounter{footnote}{0}
\section{Full $B$ Meson Reconstruction at $B$ Factories}
\label{sec:introduction}
\subsection{The Experimental Setup}
One of the biggest advantages of lepton colliders like the KEKB 
or PEP-II accelerator compared to hadron accelerators like the
Tevatron or the LHC is the precise knowledge of the initial state and 
the process of $B$ meson production. The colliding particles
are electrons and positrons. This feature allows for collisions with
well-known energy in the initial state.
As the KEKB accelerator\cite{KEKB} and the Belle detector\cite{Belle_detector} were designed to 
study $B$ meson decays, the center of mass energy of the collisions was chosen
as $\sqrt{s}=10.58$ GeV, which corresponds to the $\Upsilon (\text{4S})$ resonance.
The decay properties of this resonance are very important for the full reconstruction:

\begin{enumerate}

\item The $\Upsilon (\text{4S})$ resonance decays into a $B^{+}B^-$ or $B^{0} \bar{B^{0}}$ 
  pair respectively in over 96\% of all cases\cite{PDG} without any additional particles.

\item For the $B^{+}B^-$ or $B^{0} \bar{B^{0}}$ pairs produced in this two-body decay,
  the four-momenta are related by

  \begin{align}
   \label{equ:conservation}
    p(B_{1}) + p(B_{2}) = p(e^+) + p(e^-).
  \end{align}

\item The two $B$ mesons are almost at rest in the center of mass frame of the 
  $\Upsilon (\text{4S})$
  \begin{align}
    p_{B}^{\ast} = 380 \ \text{MeV}/c
  \end{align}
  compared to the lighter Mesons and therefore produce a spherical event topology.

\end{enumerate}
There are, however, events where no $\Upsilon (\text{4S})$, but pairs of light quarks 
($u\bar{u}$, $d\bar{d}$, $s\bar{s}$, or $c\bar{c}$) are produced. {\color{ourEm} These events form a continuum 
background to $B$ meson pair production and ideally are rejected by the analysis.}

The full reconstruction described in this paper was developed for the Belle detector\cite{Belle_detector} 
a large solid angle magnetic Spectrometer located at the KEKB collider\cite{KEKB}. 
It consists of a silicon vertex detector (SVD), a 50-layer central drift chamber (CDC), 
an array of aerogel threshold Cerenkov counters (ACC), a time-of-flight scintillation 
counter (TOF) and an electromagnetic calorimeter composed of Cs(Tl) crystals (ECL). 
All these detectors are surrounded by a superconducting solenoid, providing a 1.5 T magnetic 
field and an iron flux-return which is instrumented to detect $K^0_L$ mesons and to identify 
muons (KLM).
\subsection{The Full Reconstruction}
\begin{figure}[htbp]
  \centering
  \includegraphics[width=0.8\linewidth]{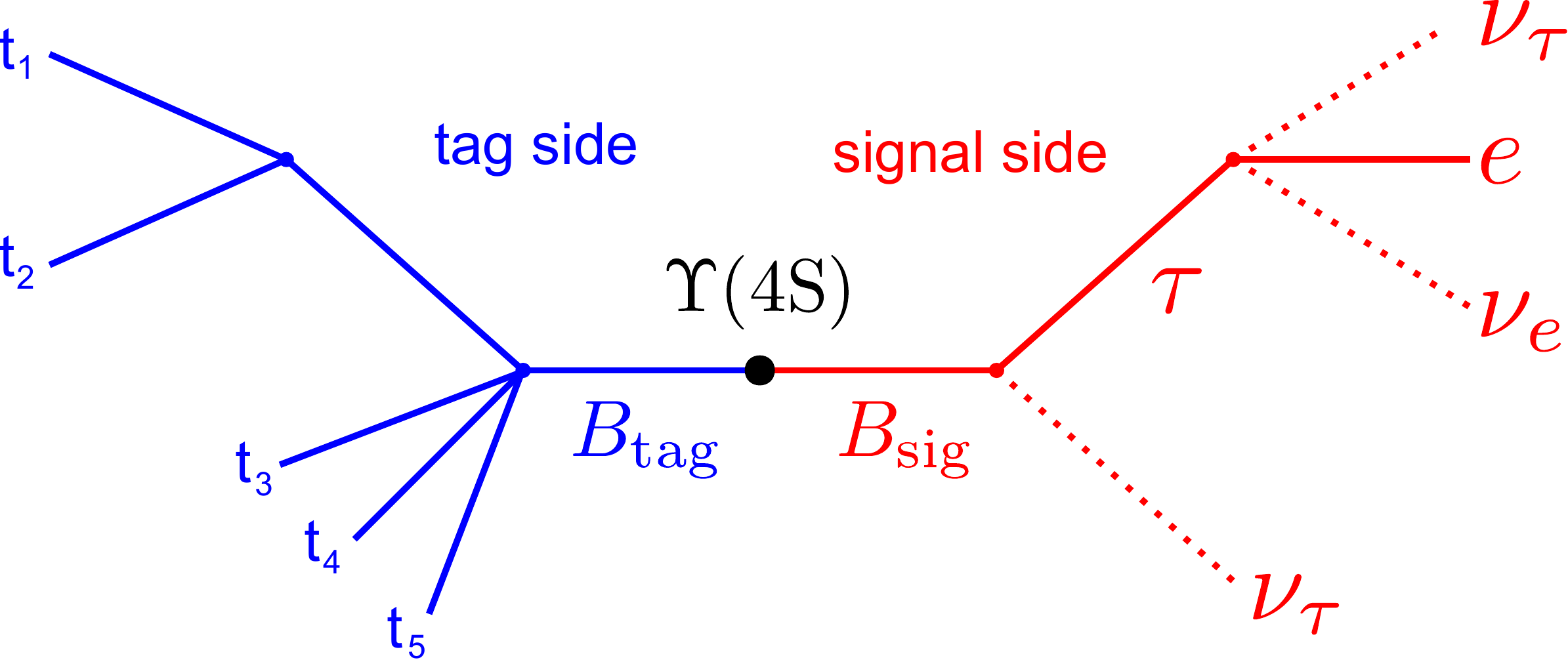}
  \caption{Exemplary fully reconstructed event. The $B_{\rm sig}$ (signal side) is the decay of physics interest, while the $B_{\rm tag}$ (tag side) is the other $B$ meson, reconstructed by the full reconstruction method.}
  \label{fig:fullrecon}
\end{figure}

The main goal and also the main difficulty of the full reconstruction is
to take any event and try to reconstruct one of the $B$ mesons in one of many 
different decay channels. Should this attempt succeed, it is 
possible to
assign all the tracks and electromagnetic clusters used in the reconstruction 
to this one $B$ meson. As it is completely reconstructed, its 4-momentum is known.
We call the fully reconstructed $B$ meson, the $B_{\rm tag}$.
After reconstruction of the tag side, it is possible to assign all the remaining
tracks and electromagnetic clusters within the detector to the other $B$ meson, which
we call the $B_{\rm sig}$ (see figure~\ref{fig:fullrecon}). This $B_{\rm sig}$ meson actually is
the object of interest for physics analyses, as explained below.

We can be sure that there are no additional particles 
produced by the $e^+ e^-$ collision within the detector, as the $\Upsilon (\text{4S})$ 
resonance decays into two $B$ mesons only.
In this two-body decay, we can obtain the momentum of the $B_{\rm sig}$ without 
any additional analysis once the $B_{\rm tag}$ is known. This 
follows by applying 4-momentum conservation as given by equation~\ref{equ:conservation}. 

This procedure might seem rather involved at first
glance, but has the benefit that it
yields information, otherwise inaccessible, about a hard or
impossible to reconstruct $B$ decay on the signal side. A prominent example
for the application of the full reconstruction is a $B$ meson decay including 
neutrinos where the decay kinematics can otherwise not be fully constrained 
or a decay with very large non-$B\bar{B}$ background.
{\color{ourEm}Many of these decays are very sensitive to small contributions from new physics
and thus it is important to adopt powerful reconstruction algorithms for them.}
Examples for the application of the full reconstruction include:

\begin{align}
 B^{+} &\to \tau^{+} \nu_{\tau}  \\
 B^{+} &\to D^{(\ast)} \tau^{+} \nu_{\tau} \\
 B^{+} &\to K^{+} \nu \nu  \\
 B^{0} &\to \nu \nu \\
 B     &\to X_u l^{+} \nu
\end{align}
One possible topology of the first decay is given in figure~\ref{fig:fullrecon},
where the $\tau$ lepton decays into an electron and two neutrinos.

The most important practical difference between the full reconstruction method and most
analyses is just the sheer number of decay channels for the
tag side. As there are several hundreds of known $B$ decay channels, 
the task of reconstructing one of the two $B$ mesons in the event cannot always succeed. Additionally, most
of those decay channels include other unstable particles, mostly $D^*$ 
and $D$ mesons, which also decay in a vast spectrum of decay channels that also have to be
reconstructed.

The quantity that has to be maximized by the full reconstruction 
method is the total $B$ reconstruction efficiency

\begin{align}
  \varepsilon_{tot}=\sum^{N}_{i} \varepsilon_{i} \cdot \mathcal{B}_{i} ,
  \label{eqn:sebi_total_B_eff}
\end{align}
where $N$ is the number of reconstructed $B$ decay channels, $\varepsilon_{i}$ is the reconstruction efficiency of the decay channel $i$ and $\mathcal{B}_{i}$ is the branching fraction of the decay channel $i$. The typical scale for $\mathcal{B}_{i}$ is $10^{-3}$ to $10^{-5}$ and typically $\varepsilon_{i}$ is of the order of $10\%$. As the $\mathcal{B}_{i}$ is fixed by nature, we can maximize $\varepsilon_{tot}$ only by increasing $\varepsilon_{i}$ and the number of reconstructed decay channels $N$. In order to increase $\varepsilon_{i}$, multivariate techniques are used (see chapter~\ref{chap:multivariate_techniques}). The main challenge is to keep track of all the used variables in these multivariate methods, particularly because we want to reconstruct as many decay channels as possible. 
{\color{ourEm}For this we had to develop a software framework which gives us the possibility to automatically manage hundreds of decay channels with extensive usage of multivariate methods. The automatic handling of many steps allows to minimize human errors.}

\section{Multivariate Techniques}
\label{chap:multivariate_techniques}
A common technique to achieve more sophisticated selections is to combine 
all significant variables available into a single scalar variable, for
example a likelihood ratio, and to perform a cut on this new variable.
These multivariate techniques are in principle capable of taking 
correlations of the variables into account. The application of these
techniques can, however, be rather involved. Simplified models can 
deliver quite good results when correlations between the different 
variables are small.

Another example of a multivariate technique is the NeuroBayes
package~\cite{neurobayes} that was used extensively for the new full reconstruction tool.
The idea of the NeuroBayes package is to pass all of the
relevant variables, through a pre\-processing algorithm, 
to a neural network. 
For a classification task, to decide if a candidate is signal or background, 
the network maps the input variables to a single output variable
while taking into account the correlations of the input variables. {\color{ourEm} An example of the separation power of this output variable for one of the classification task used can be seen in figure~\ref{fig:nbout}.}

\subsection{NeuroBayes Output as a Probability}
\label{sec:nb_probability}
\begin{figure}[htbp]
  \centering
  \subfigure[]
  {
    \includegraphics[width=0.45\linewidth, angle=0]{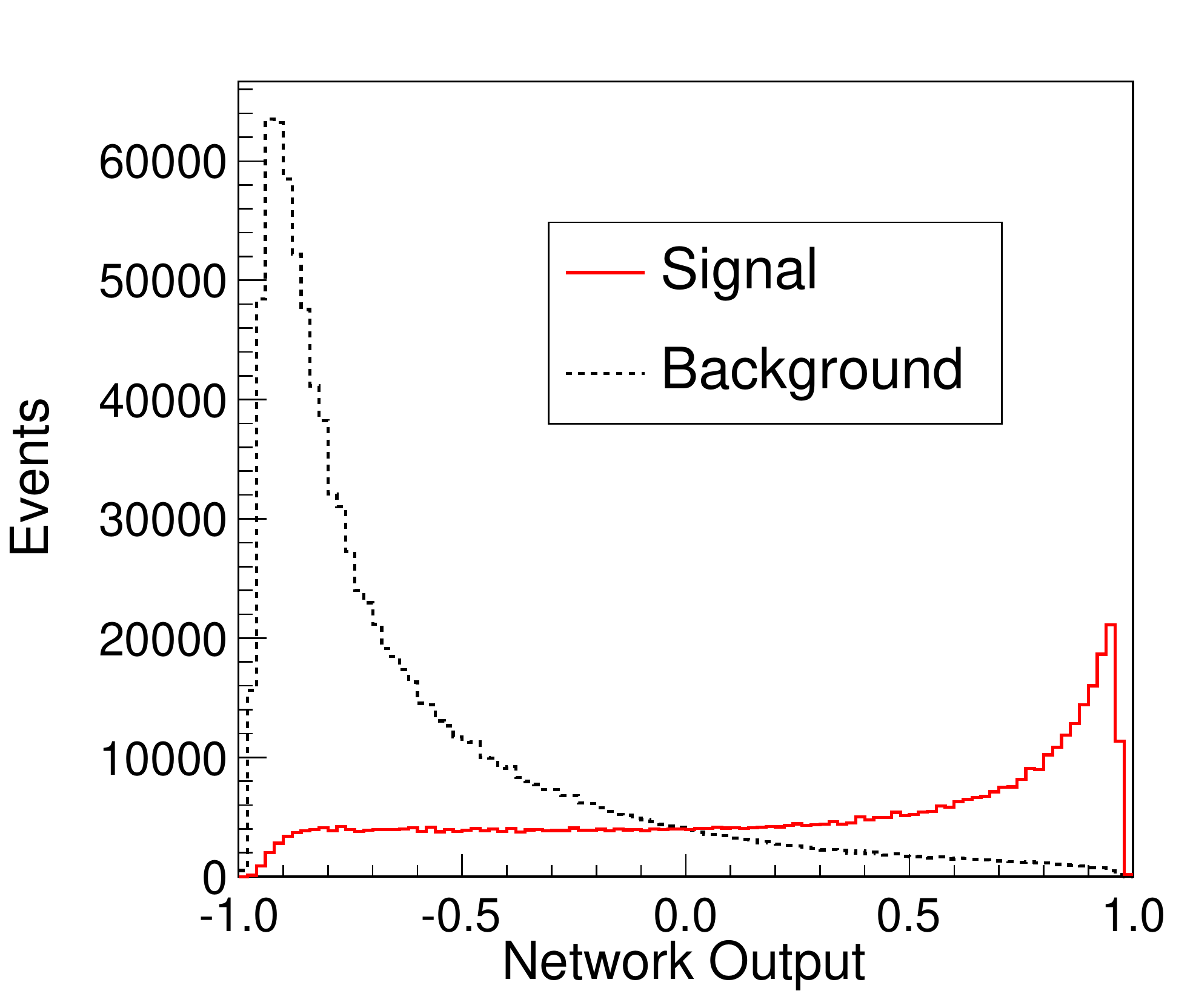}
  \label{fig:nbout}
  }
  \subfigure[]
  {
    \includegraphics[width=0.45\linewidth, angle=0]{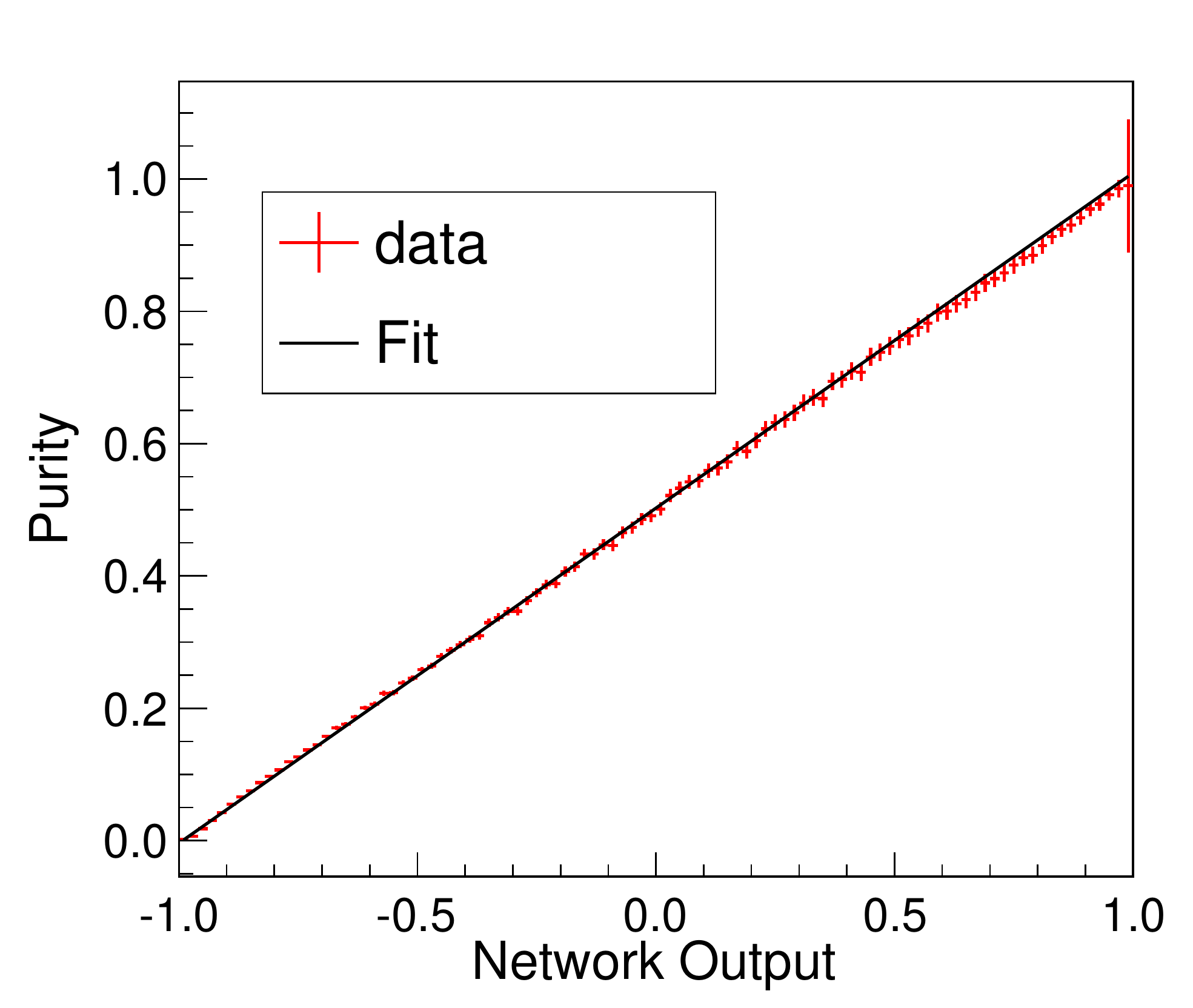}
  \label{fig:purity_nbout}
  }
  \caption[Optional caption for list of figures]{~\subref{fig:nbout}: The distribution of the NeuroBayes output for signal (red) and background (black) for an exemplary classification task of $\pi^0$ candidates. ~\subref{fig:purity_nbout}: The purity, obtained from the network output distributions shown in Fig.~\subref{fig:nbout},
is a linear function of the NeuroBayes output.}
\end{figure}

As shown in figure~\ref{fig:purity_nbout}, the purity, defined as the number of signal events divided by the total number of events in a network output bin, is a linear function of the NeuroBayes output.
This indicates that the produced output is a good measure of probability for the candidate to be signal.

If a NeuroBayes training is performed with the same signal to background ratio
as found on data, the output of the classification can directly be interpreted
as a Bayesian probability for signal.
While it would be better to train
the neural network with the same signal to background ratio as expected on data, 
it is sometimes not possible. If, for example, the desired signal is very rare in 
nature, a training would not learn to distinguish the few signal events
from the millions of background events, but rather try to learn something
from statistical fluctuations of the background that swamp the signal 
and therefore also dominate the loss function that is minimized during the network training. 
Therefore, a training with a higher 
signal fraction is the only way, in which the selection of such rare signals can be optimized.
On the other hand, if we artificially increase the signal to background ratio, the network output cannot be interpreted as a Bayesian probability any more on the real dataset, because the a priori probabilities of being signal or background differ from the training dataset. Nevertheless, one can correct the network output in a way that is interpretable as a probability again. For this, we need to know the signal to background ratio in the training dataset and in the dataset where the network should give the prediction. 
To calculate this correction, we need  Bayes' theorem, which is defined for two types of events, $X$ and $Y$, as

\begin{align}\label{bayestheorem}
P(X\vert Y) &= \frac{P(Y\vert X)P(X)}{P(Y)} \;.
\end{align}
{\color{ourEm}For our purposes, it is preferable to use Bayes' theorem in terms of the likelihood ratio
\begin{align}
\varLambda (Y\vert X) = \frac{P(Y\vert X)}{P(Y\vert \neg X)} \;,
\end{align}
which leads to prior odds of
\begin{align}\label{priorodds}
O(X) = \frac{P(X)}{P(\neg X)} \;, 
\end{align}
and posterior odds given by
\begin{align}\label{bayestheorem_likelihood}
O(X\vert Y) &= O(X) \cdot \varLambda (Y\vert X) \;,
\end{align}
In our example $X$ and $\neg X$ are signal events ($S$) and background events ($B$) and $Y$ is the output ($o_t$) from a network trained with the training dataset (denoted with the subscript $t$). The likelihood ratio is
\begin{align}\label{likelihoodratio}
\varLambda (Y\vert X) = \frac{P(o_t\vert S)}{P(o_t\vert B)} \;,
\end{align}
where $P(o_t\vert S)$ is the likelihood to get a network output, $o_t$, given a signal event $S$ and $P(o_t\vert B)$ is the same for a background event.
Given a network output $o_{t}$ the conditional probability of being a signal event $S$, is
\begin{equation}\label{oiistgleichbedintew}
 o_{t}=P_{t}(S\vert o_{t}) \;,
\end{equation}
and the corresponding probability of being a background event $B$ is given by
\begin{equation}
(1-o_{t})=P_{t}(B\vert o_{t}) \;.
\end{equation}
By applying Bayes theorem as follows
\begin{align}
\frac{P_t(S\vert o_t)}{P_t(B\vert o_t)}&=\frac{P_t(S)}{P_t(B)} \cdot \varLambda (o_t\vert S)  \;
\end{align}
we can write the likelihood ratio as
\begin{align}
\varLambda (o_t\vert S) &= \frac{P(o_t\vert S)}{P(o_t\vert B)}\nonumber\\
		      &= \frac{o_t}{1-o_t}\cdot \frac{P_t(B)}{P_t(S)}\;.
\end{align}
}
This likelihood ratio does not depend on the signal to background ratio because it only contains measured information of one given event. We can now calculate, for any other signal to background ratio in the prediction dataset (denoted with the subscript $p$), the posterior odds with Bayes theorem:
%
\begin{align}
\frac{P_p(S\vert o_p)}{P_p(B\vert o_p)}&=\frac{P_p(S)}{P_p(B)} \cdot \varLambda (o_t\vert S)  \;.
\end{align}
Because the transformed probability $o_p$ has to satisfy

\begin{align}
\frac{P_p(S\vert o_p)}{P_p(B\vert o_p)}	&=\frac{o_p}{1-o_p} 
\end{align}
to be the correct probability, we get:

\begin{align}\label{umrechnungnetout}
o_p&= \frac{1 }{1+(\frac{1}{o_{t}}-1)\frac{P_{p}(B)}{P_{p}(S)}\frac{P_{t}(S)}{P_{t}(B)}} \;.
\end{align}
This formula is used in the full reconstruction algorithm described in the next section
to calculate the signal probability for modes with low purity so that the signal fraction
had to be increased for the network training.

\section{Selection and Reconstruction}
\label{sec:selection}
\begin{figure}[htbp]
  \centering
  \includegraphics[width=0.8\linewidth]{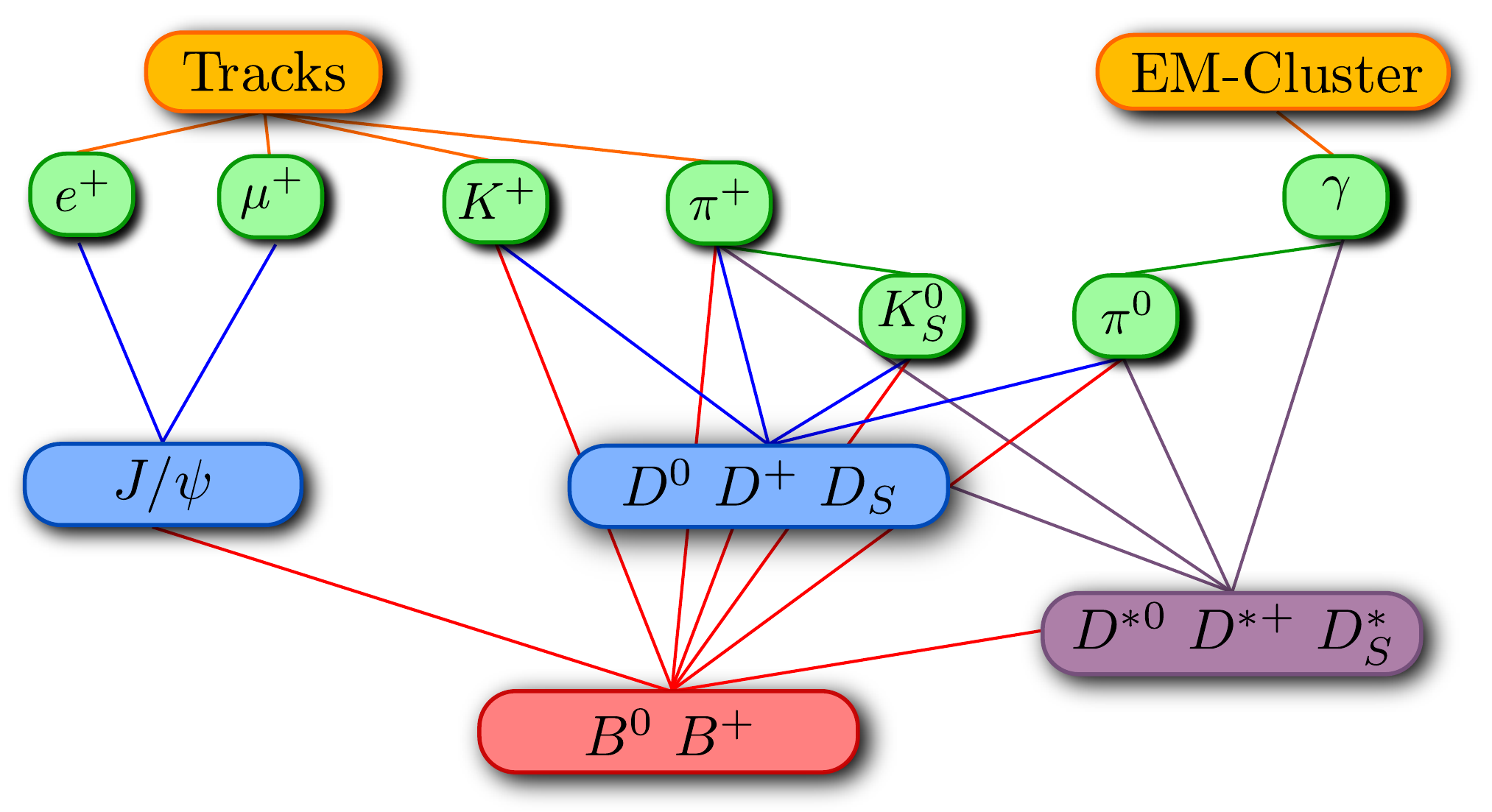}
  \caption{The 4 stages of the full reconstruction}
  \label{fig:the_4_stages}
\end{figure}

In order to reconstruct as many $B$ meson decays as possible,
it is not possible to take care of the thousands of exclusive decay channels individually. Instead a hierarchical 
approach was chosen. We divide the reconstruction into
4 stages, as shown in table~\ref{tab:stages} and illustrated in 
figure~\ref{fig:the_4_stages}.

\begin{table}
  \centering
  \begin{tabular}{|c|c|}
    \hline
    stage & particles \\ \hline \hline
    1 & tracks, $K_S$, $\gamma$, $\pi^0$          \\ \hline
    2 & $D_{(s)}^{\pm}$, $D^0$, and $J/\psi$ mesons  \\ \hline
    3 & $D_{(s)}^{*\pm}$ and $D^{*0}$ mesons        \\ \hline
    4 & $B^{\pm}$ and $B^{0}$ mesons               \\ \hline
  \end{tabular}
  \caption{The 4 stages of the hierarchical system}
  \label{tab:stages}
\end{table}
One aim of the full reconstruction is to achieve high efficiency. This could in 
theory be done by always reconstructing every possible candidate at all stages in an event and
then finally taking the best $B$ meson candidate. In practice however, the computing power needed 
to pursue this maximum efficiency strategy is not available and it is necessary
to perform cuts during the selection and reconstruction process. A main principle 
of this ansatz is to calculate the signal probabilities at each stage of
the hierarchical system, while cuts on these probabilities occur only at a later stage.

\subsection{Data Samples}
For the training we used a Monte-Carlo generated data sample with a full detector simulation 
based on GEANT\cite{GEANT}. It includes $e^+e^-$ annihilation to non-$b$ quarks ($u$, $d$, $s$, $c$)
 generated with PYTHIA\cite{PYTHIA} and to the $\Upsilon(4S)$ resonance generated with the EvtGen 
package \cite{Lange:2001}. {\color{ourEm}The produced mesons then decay inclusively to any possible final state governed by the  $b\rightarrow c$ transition.}

\subsection{The First Stage}
In the \emph{first stage}, NeuroBayes networks are trained on Monte-Carlo samples for charged, 
long-lived particle type hypotheses (kaon, pion, electron, muon) for the measured charged tracks 
and for the photon hypothesis for each cluster in the electromagnetic calorimeter not matched 
geometrically to a charged track. Neutral pion candidates are formed out of two electromagnetic clusters
whose invariant mass lies within the window $115$ MeV$/c^2<M(\pi^0)<153$ MeV$/c^2$. Additionally,
the energy of the photons that are used to construct the $\pi^0$ candidate has to lie above $30$ MeV. 
Candidate $K_{S}$ particles are formed from two charged tracks whose invariant mass lies within $30$ MeV of
the nominal $K_S^{0}$ mass.
Only very loose preselection criteria on the impact parameter of all tracks and the
particle identification variable \cite{Belle_detector} for $K^{+}$ candidates are applied. {\color{ourEm}As an example in the decay
$D^{0} \to K^{-} \pi^{+}$, a signal efficiency of $96\%$ and a background reduction factor of
$3.5$ could be observed.}
After these pre-cuts, NeuroBayes networks are trained for all particle hypotheses.
As an input for the trainings of the charged particles, measurements of the time-of-flight, 
the energy loss in the CDC and Cherenkov light in the ACC are used. For the photon hypothesis, 
several variables to describe the shower shape in the calorimeter are used.

\subsection{The Second Stage}
In the \emph{second stage}, combinations of two to five candidates from the first stage were used to reconstruct
$D^{\pm}$, $D^{0}$, $D_s^{\pm}$ and $J/\psi$ mesons. A list of the decay channels used for the
reconstruction of these mesons and their respective branching fractions can be found in table~\ref{table:channels_s2}.

\begin{table}[htbp]
  \centering
  \begin{tabular}{|llr||llr|}
    \hline
    \multicolumn{3}{|c||}{$D^0$} & \multicolumn{3}{c|}{$D^+$} \\ \hline \hline
    \multicolumn{2}{|c}{mode}& \multicolumn{1}{c||}{BR} &\multicolumn{2}{c}{mode} & \multicolumn{1}{c|}{BR} \\ \hline

    $D^{0} \ \ \rightarrow$ & $K^{-} \pi^{+}$                   & $3.89\%$ &  $D^{+} \ \ \rightarrow$ &$K^{-} \pi^{+} \pi^{+}$                 &$9.40\%$    \\ \hline 
    $D^{0} \ \ \rightarrow$ & $K^{-} \pi^{+} \pi^{+} \pi^{-}$   & $8.09\%$&   $D^{+} \ \ \rightarrow$ & $K_{S}^{0} \pi^{+}$                    &$1.49\%$     \\ \hline 
    $D^{0} \ \ \rightarrow$ & $K^{-} \pi^{+} \pi^{0}$           &$13.90\%$ &   $D^{+} \ \ \rightarrow$ &   $K_{S}^{0} \pi^{+} \pi^{0}$          &$6.90\%$     \\ \hline 
    $D^{0} \ \ \rightarrow$ & $\pi^{+} \pi^{-}$                 &$0.14\%$ &   $D^{+} \ \ \rightarrow$ &   $K^{-} \pi^{+} \pi^{+} \pi^{0}$      &$6.08\%$        \\ \hline 
    $D^{0} \ \ \rightarrow$ & $\pi^{+} \pi^{-} \pi^{0}$         &$1.44\%$ &   $D^{+} \ \ \rightarrow$ &   $K_{S}^{0} \pi^{+} \pi^{+} \pi^{-}$  & $3.10\%$       \\ \hline 
    $D^{0} \ \ \rightarrow$ & $K_{S}^{0} \pi^{0}$               &$1.22\%$ &   $D^{+} \ \ \rightarrow$ &   $K^{+} K^{-} \pi^{+}$                & $0.98\%$       \\ \hline 
    $D^{0} \ \ \rightarrow$ & $K_{S}^{0} \pi^{+} \pi^{-}$       &$2.94\%$ &   $D^{+} \ \ \rightarrow$ &    $K^{+} K^{-} \pi^{+} \pi^{0}$        &$1.50\%$       \\ \hline   
    $D^{0} \ \ \rightarrow$ & $K_{S}^{0} \pi^{+} \pi^{-} \pi^{0}$&$5.40\%$&                           &                                        &          \\ \hline               
    $D^{0} \ \ \rightarrow$ & $K^{+} K^{-}$                     &$0.39\%$ &                          &                                         &         \\ \hline
    $D^{0} \ \ \rightarrow$ & $K^{+} K^{-} K_{S}^{0}$           &$0.47\%$ &                          &                                         &          \\ \hline

%
    \hline \hline
    \multicolumn{3}{|c||}{$D_s$} & \multicolumn{3}{c|}{$J/\psi$} \\ \hline \hline
    \multicolumn{2}{|c}{mode}& \multicolumn{1}{c||}{BR} &\multicolumn{2}{c}{mode} & \multicolumn{1}{c|}{BR} \\ \hline

 $D_{s}^{+} \ \ \rightarrow$ & $K^{+} K_{S}^{0}$      &$1.49\%$&$J/\psi \ \ \rightarrow$ & $e^{-} e^{+}$ & $5.94\%$ \\ \hline
     $D_{S}^{+} \ \ \rightarrow$ & $K^{+} \pi^{+} \pi^{-}$ &$0.69\%$&$J/\psi \ \ \rightarrow$ & $\mu^{-} \mu^{+}$& $5.93\%$ \\ \hline
      $D_{s}^{+} \ \ \rightarrow$ & $K^{+} K^{-} \pi^{+}$  &$5.50\%$&&& \\ \hline
     $D_{s}^{+} \ \ \rightarrow$ & $K^{+} K^{-} \pi^{+} \pi^{0}$   &$5.60\%$&&& \\ \hline        
     $D_{s}^{+} \ \ \rightarrow$ & $K^{+} K_{S}^{0} \pi^{+} \pi^{-}$&$0.96\%$&&& \\ \hline       
     $D_{S}^{+} \ \ \rightarrow$ & $K^{-} K_{S}^{0} \pi^{+} \pi^{+}$ &$1.64\%$&&& \\ \hline      
      $D_{S}^{+} \ \ \rightarrow$ & $K^{+} K^{-} \pi^{+} \pi^{+} \pi^{-}$ &$0.88\%$&&& \\ \hline 
  $D_{s}^{+} \ \ \rightarrow$ & $\pi^{+} \pi^{+} \pi^{-}$      &$1.10\%$&&& \\ \hline            
  \end{tabular}
  \caption{Stage 2 - Reconstructed $D$ and $J/\psi$ modes. Branching ratios are from Ref.~\cite{PDG}.}
  \label{table:channels_s2}
\end{table}
{\color{ourEm}As these trainings were performed on inclusive simulated samples, a large fraction of
the true $D$ mesons did not come from $B$ meson decays.} Since only the $D$ 
mesons from $B$ decays are of interest, a momentum cut in the $\Upsilon(4S)$ rest
frame was performed: 

\begin{align}
  p^{\ast}(D) < 2.6 \ \text{ GeV}/c
\end{align}
This cut excludes the majority of $D$ mesons not stemming from $B$ decays, i.e. from $c\bar{c}$-fragmentation.

\subsubsection{Selection Criteria}
%
{\color{ourEm}
In order to retain reasonable computing time, soft pre-cuts are applied at this stage. We define the quantity
\begin{align}
  \text{NB}_{out,prod} &= \prod_{i}^{N} \text{NB}_{out,i}  \ ,
\label{eqn:prod_child_nb}
\end{align}
where $N$ is the number of daughters in a given decay and $\text{NB}_{out,i}$ is the neural network output of $i$-th daughter, which we use to suppress obvious background.
}

\begin{figure}[htbp]
  \centering
  \includegraphics[width=\linewidth]{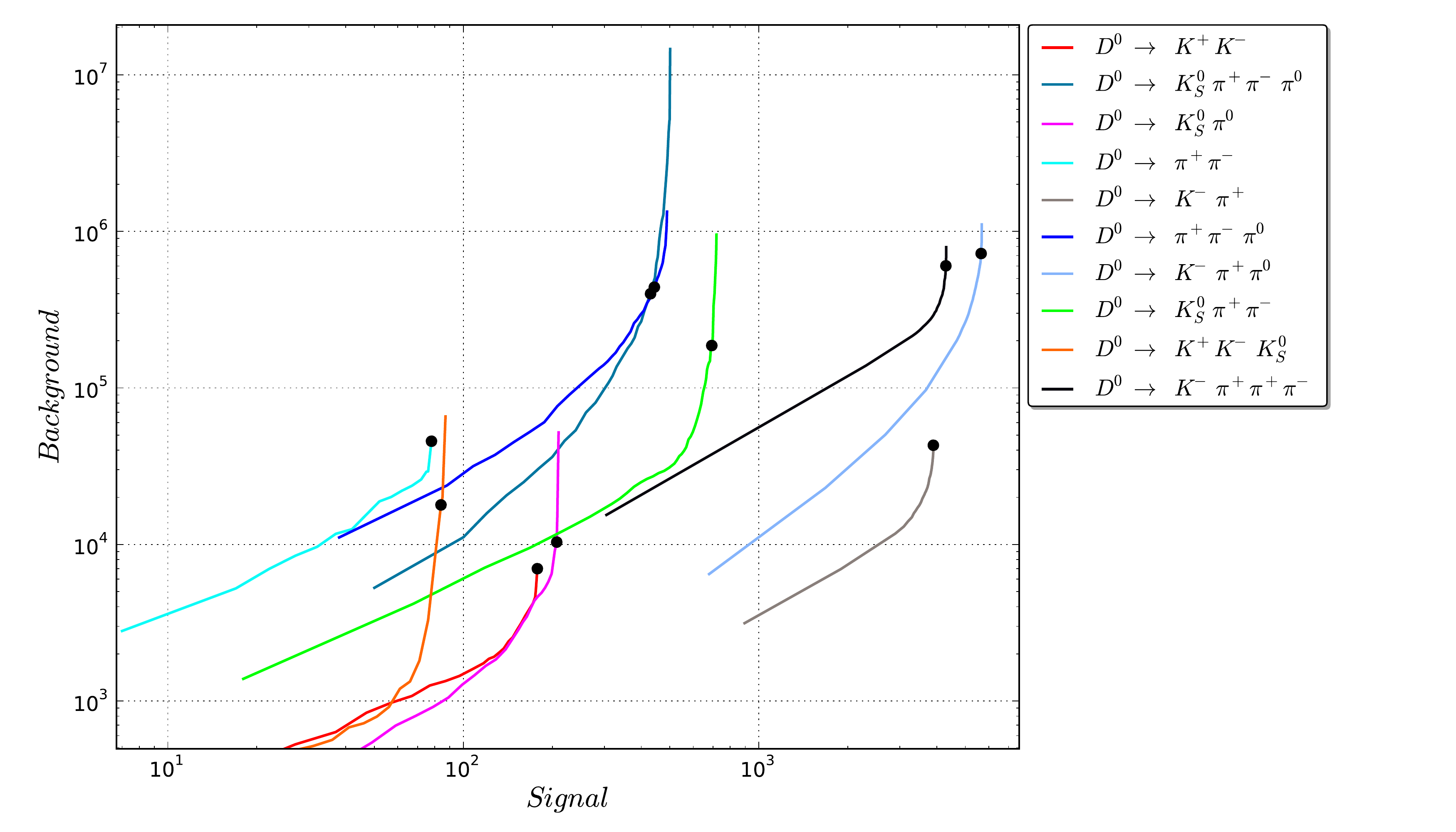}
  \caption{The signal-background plots for the $D^{0}$ cut determination. The black dots are our cutting points and all the lines have the same slope in these points. (For colored lines, see the online version of this paper)}
  \label{fig:prodchildnb_cut_D0}
\end{figure}

The cuts for all decay modes of a particle type were determined simultaneously
to optimally use the CPU resources.
To explain the determination of the cuts, let us focus on $D^{0}$ mesons:
The cuts were determined for all $D^{0}$ modes simultaneously. It was required
that the additional amount of background that would have to be taken into the sample
to gain one additional signal event was the same for all $D^{0}$ modes. This means
that very clean channels will get a very soft cut and at the same time, more complicated 
channels will get slightly harder cuts so that the consumed computing power is minimized. To determine these cuts,
for each $D^{0}$ mode the number of signal events in the sample was plotted against the 
number of background events for the different possible cuts on the product of the 
NeuroBayes outputs of the children.  If we now look at the slopes of the tangents of 
these different plots, the same tangent slope indicates the same additional number of 
background events for one additional signal. 
{\color{ourEm}The cut is set at that value where this condition is met.}
Figure~\ref{fig:prodchildnb_cut_D0} shows a possible choice for the slope.

The steeper this slope is, the higher the efficiency is, but also the higher computational effort is 
needed in these modes. The final choice of the exact value of each slope is obviously an
arbitrary matter. Our decision for the slopes of $D^{\pm}$, $D^{0}$, $D_s^{\pm}$ and also those of $D^{*}$ 
modes in stage 3 were made from the point of view of combining these particles to a $B$ meson and then 
getting on average much less than one candidate per event.
All of the remaining candidates for $D^{\pm}$, $D^{0}$, $D_s^{\pm}$ and $J/\psi$ mesons
were again classified using NeuroBayes. 
The networks comprise a large number of variables. The variables with the largest separation power are the product 
of the NeuroBayes outputs of the children, the invariant mass of children pairs and the angle between them,
the angle between the momentum of the $D$ meson and the line connecting the $D$ decay vertex to the interaction point
and the significance of the distance of the $D$ meson decay vertex to the interaction point.

\begin{figure}[htbp]
  \centering
  \subfigure[\ $D^{+} \to K_s^{0} \ \pi^{+}$ - with no network cut]
  {
    \includegraphics[width=0.40\linewidth, angle=0]{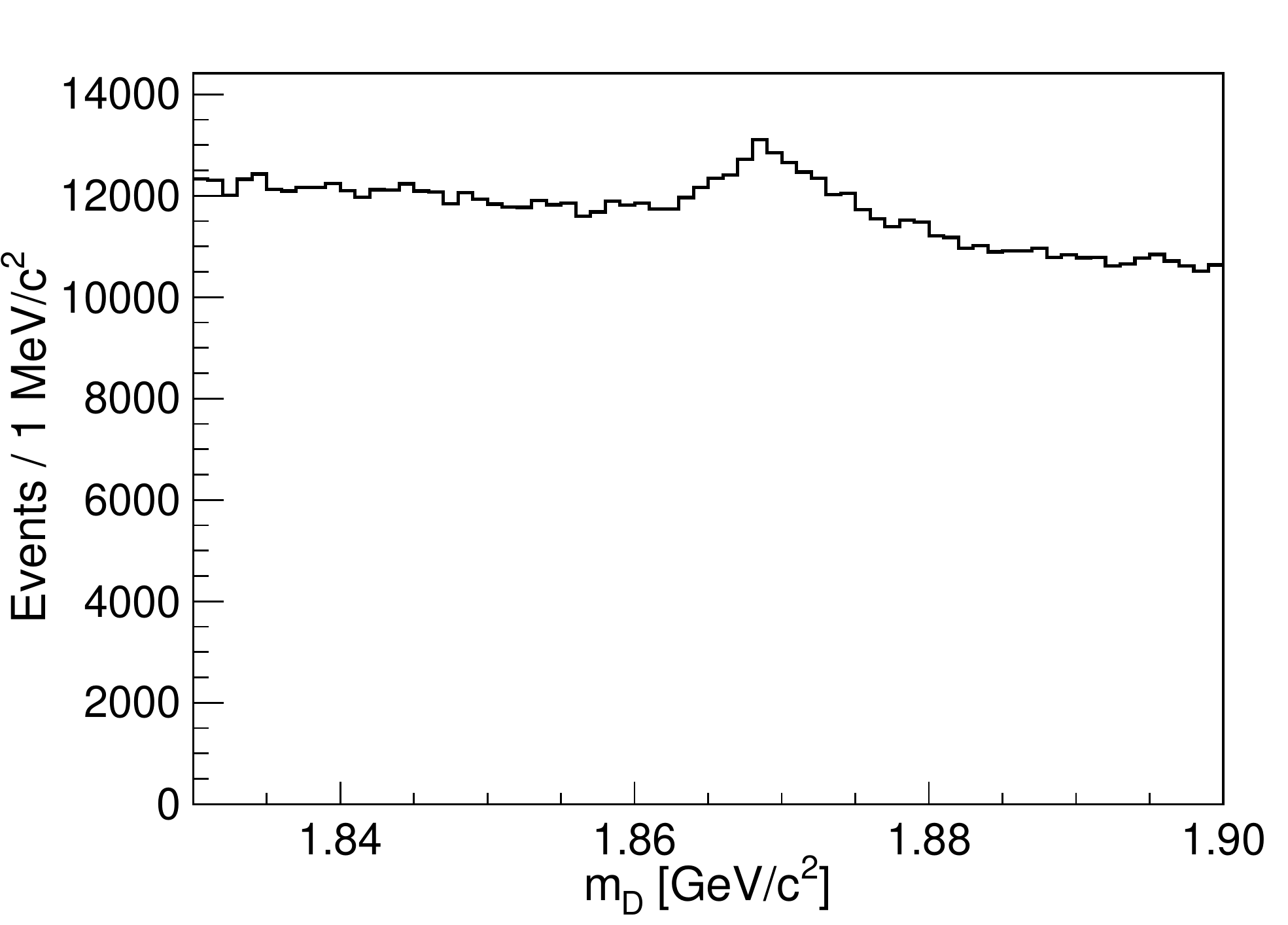}
    \label{fig:dp_kspi}
  }
  \subfigure[\ $D^{+} \to K_s^{0} \ \pi^{+}$ - with an arbitrary network cut]
  {
    \includegraphics[width=0.40\linewidth, angle=0]{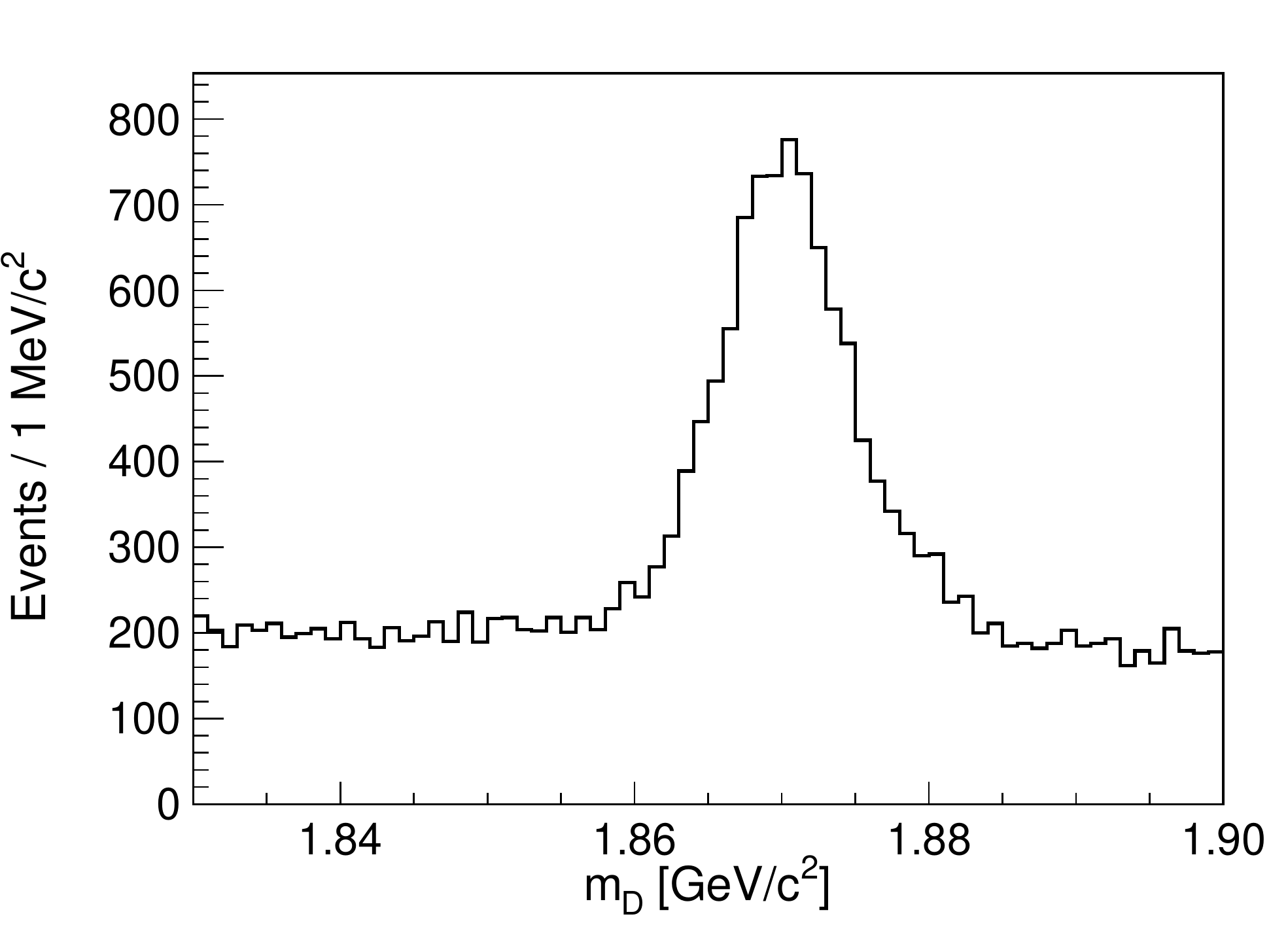}
    \label{fig:dp_kspi_2}
  }
  \subfigure[\ $D^{+} \to K^{-} \ \pi^{+} \ \pi^{+}$ - with no network cut]
  {
    \includegraphics[width=0.40\linewidth, angle=0]{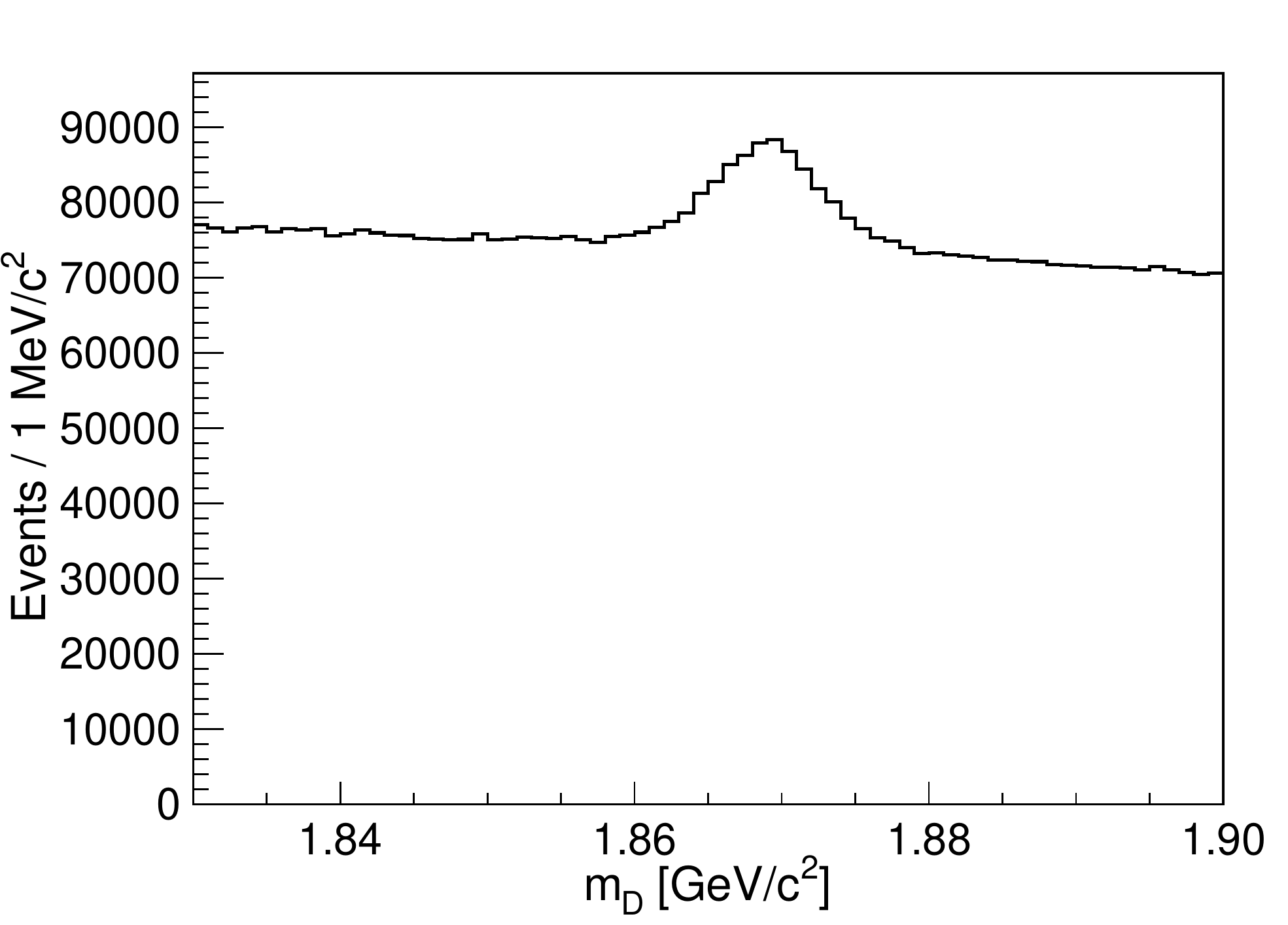}
    \label{fig:dp_kpipi}
  }
  \subfigure[\  $D^{+} \to K^{-} \ \pi^{+} \ \pi^{+}$ - with an arbitrary network cut]
  {
    \includegraphics[width=0.40\linewidth, angle=0]{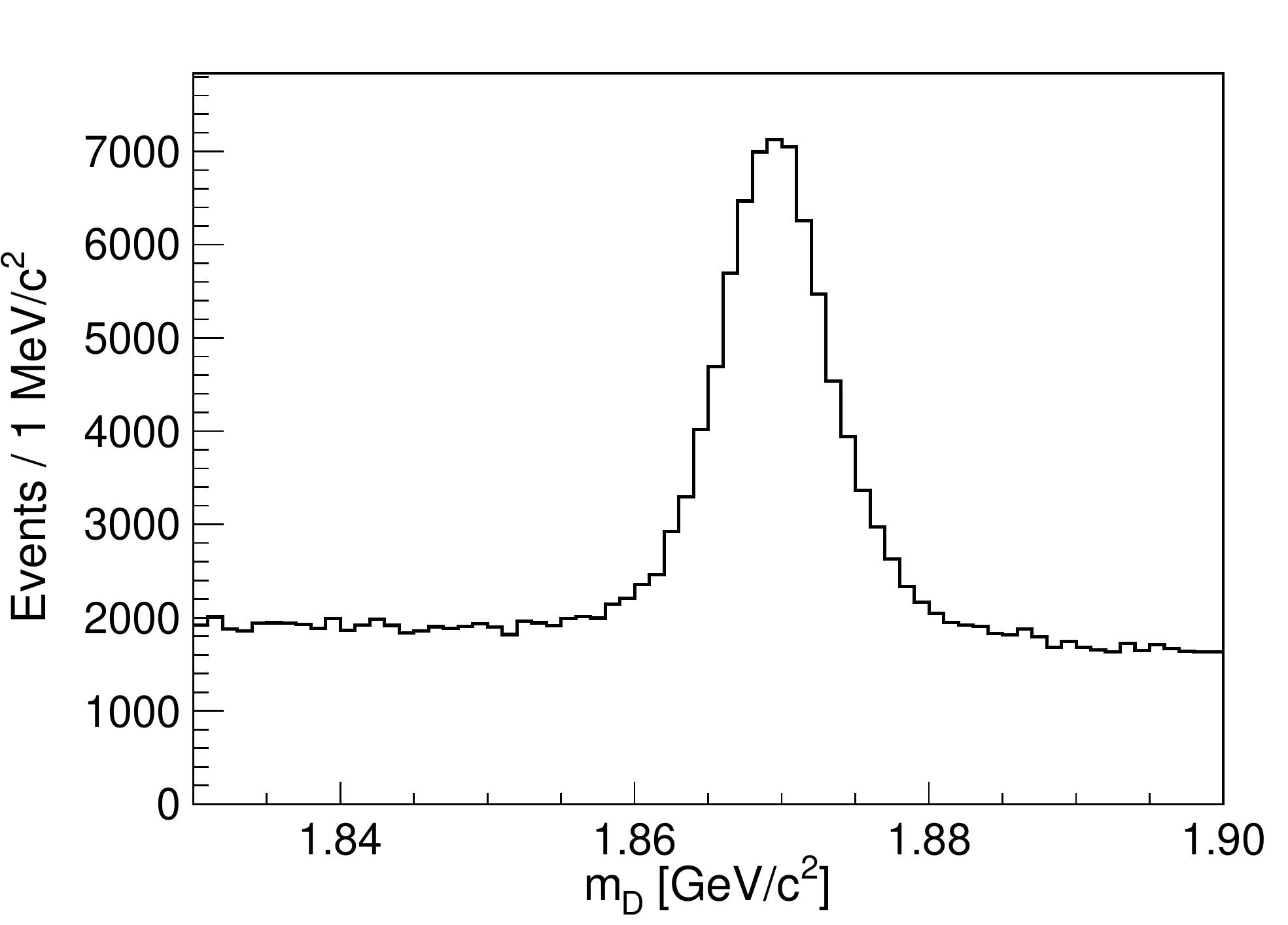}
    \label{fig:dp_kpipi_2}
  }
  \caption[Optional caption for list of figures]{Intermediate results: $D^{+}$ mesons}
  \label{fig:dp_intermediate}
\end{figure}
\begin{figure}[htbp]
  \centering
  \subfigure[\ $D^{0} \to K^{-} \ \pi^{+} \ \pi^{0}$ - with no network cut]
  {
    \includegraphics[width=0.40\linewidth, angle=0]{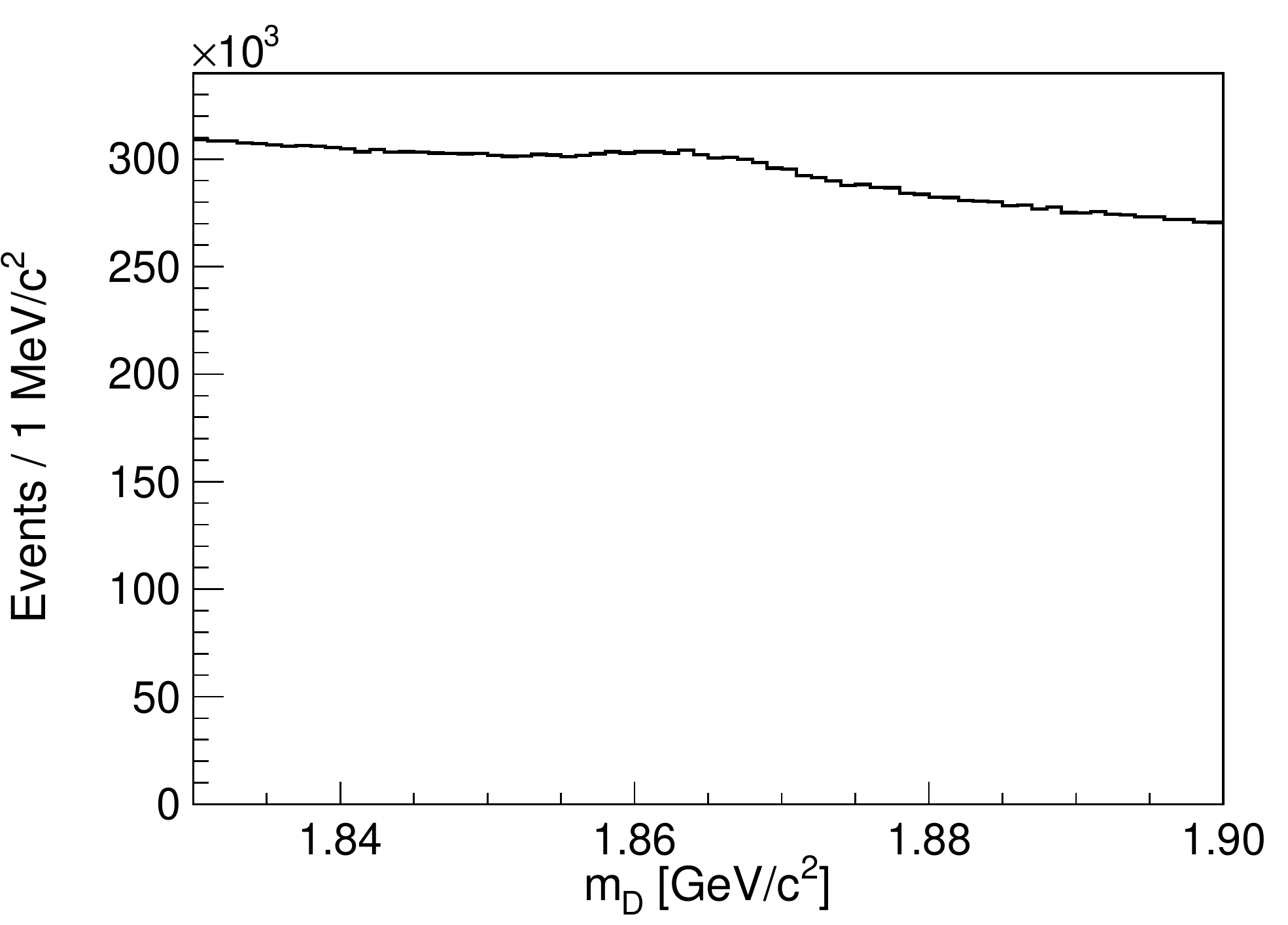}
    \label{fig:d0_kpipi}
  }
  \subfigure[\ $D^{0} \to K^{-} \ \pi^{+} \ \pi^{0}$ - with an arbitrary network cut]
  {
    \includegraphics[width=0.40\linewidth, angle=0]{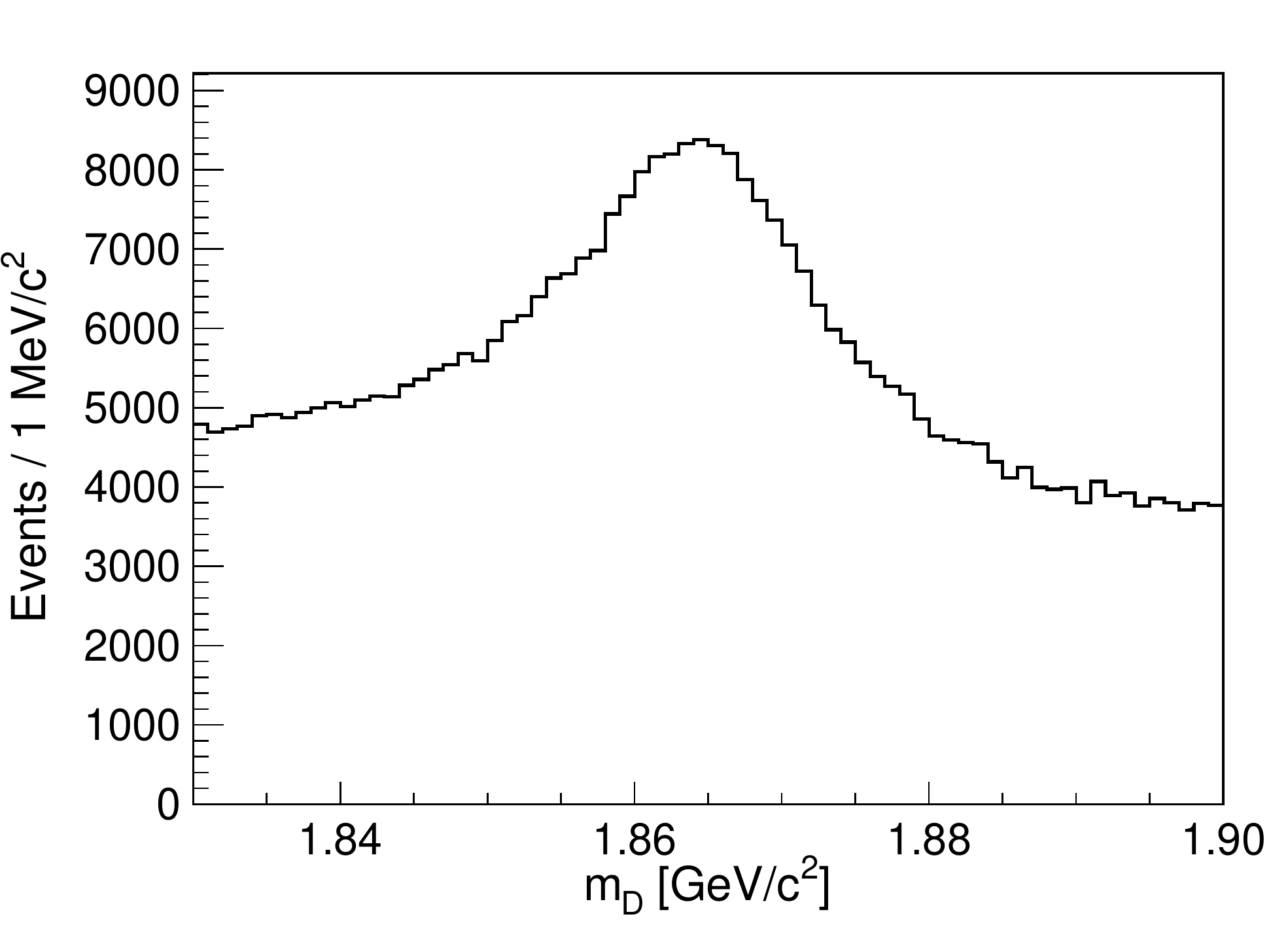}
    \label{fig:d0_kpipi_2}
  }
  \subfigure[\ $D^{0} \to  K_S^{0} \ \pi^{+} \ \pi^{+} \ \pi^{0}$ - with no network cut]
  {
    \includegraphics[width=0.40\linewidth, angle=0]{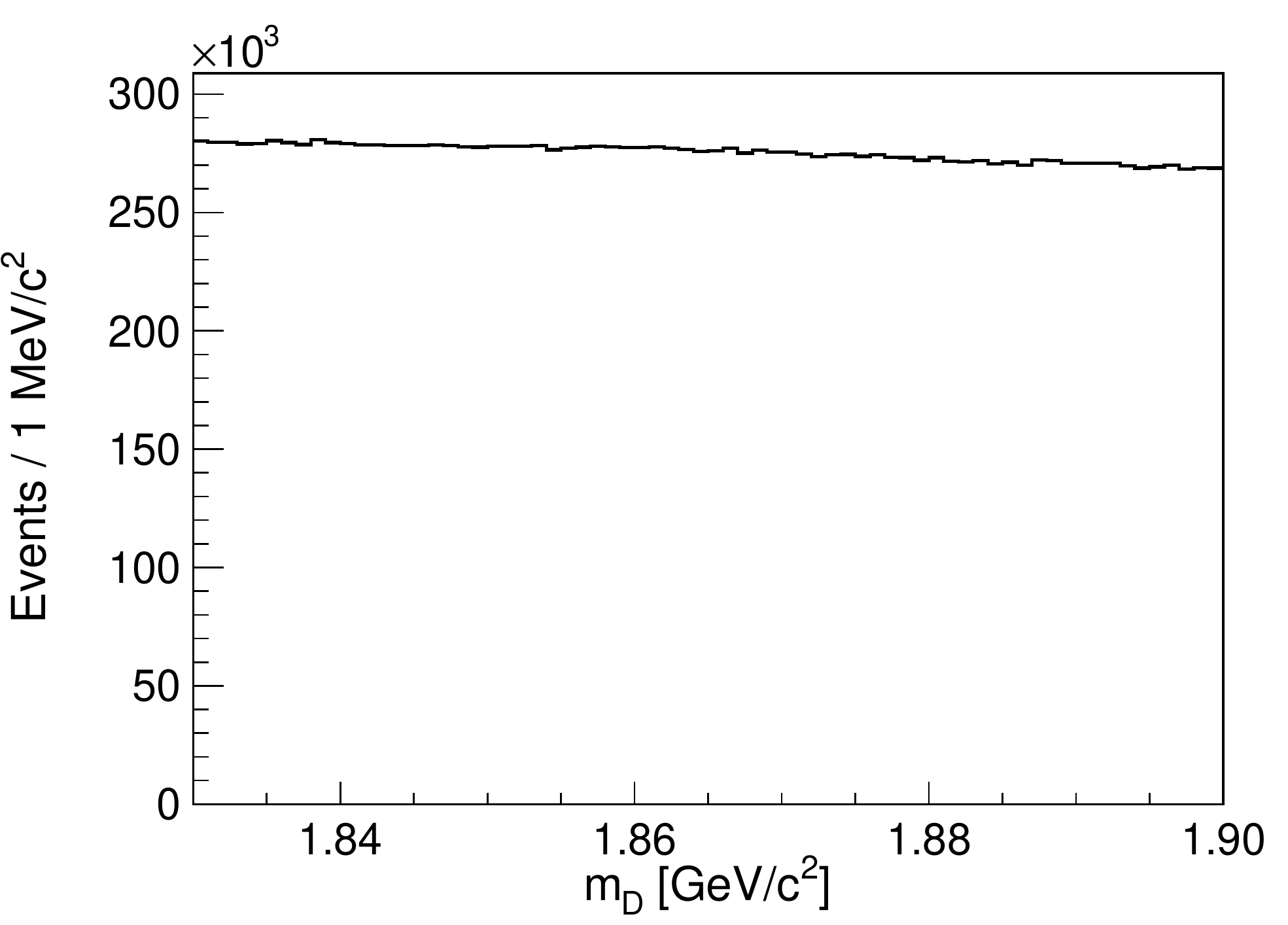}
    \label{fig:d0_k3pi}
  }
  \subfigure[\  $D^{0} \to  K_S^{0} \ \pi^{+} \ \pi^{+} \ \pi^{0}$ - with an arbitrary network cut]
  {
    \includegraphics[width=0.40\linewidth, angle=0]{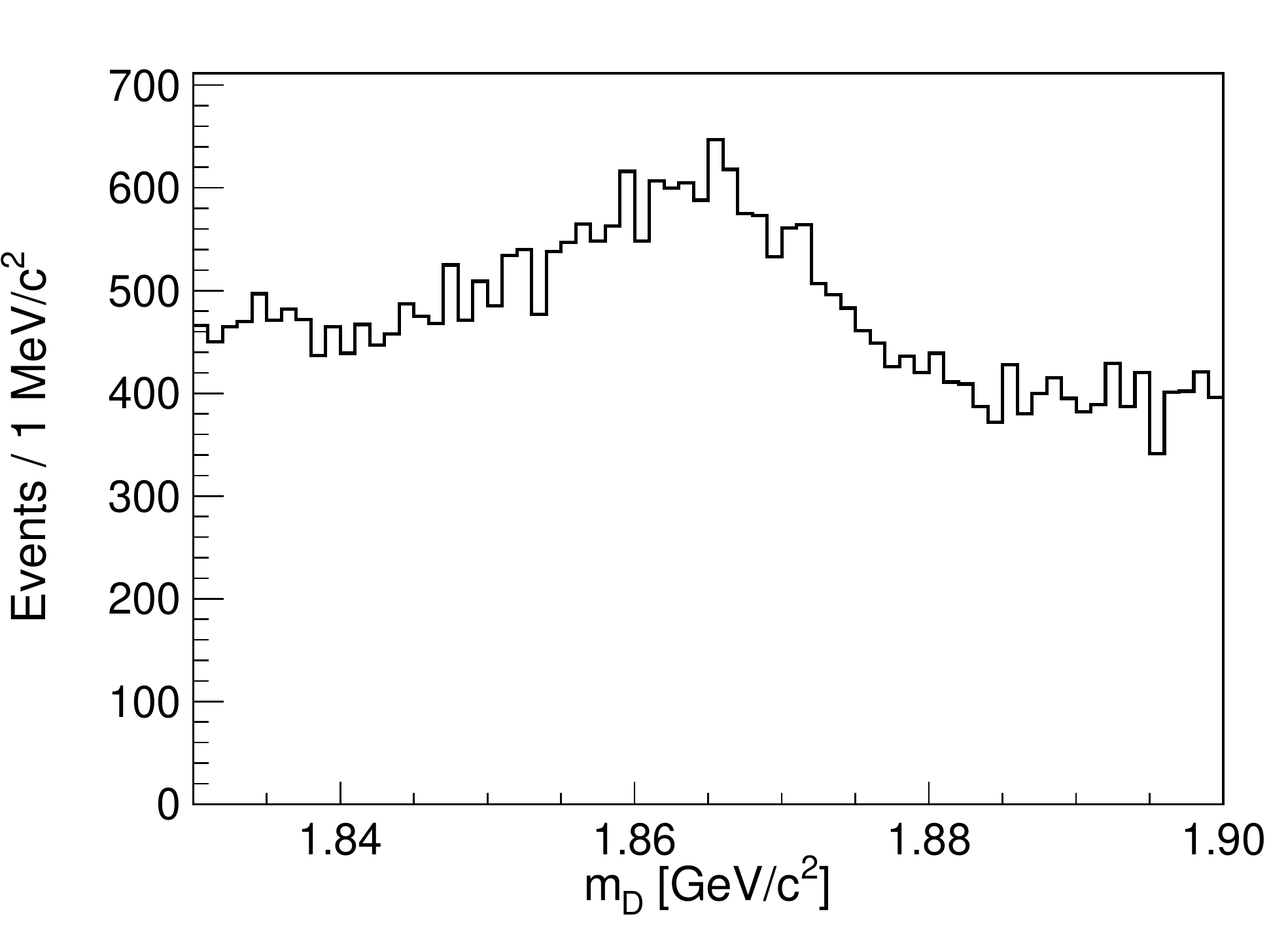}
    \label{fig:d0_k3pi_2}
  }
  \caption[Optional caption for list of figures]{Intermediate results: $D^{0}$ mesons}
  \label{fig:d0_intermediate}
\end{figure}

Special attention
was paid to not include any mass-dependent variable in these trainings, so that, if necessary, 
a check of our $D$ and $J/\psi$-meson sample could be performed by looking at the 
unbiased mass distributions.
Examples of few intermediate results for a few stage 2 channels are shown in figures~\ref{fig:dp_intermediate}
and~\ref{fig:d0_intermediate}. 

In the hierarchical system, it is very important that all the outputs of the 
NeuroBayes trainings actually represent their signal probabilities, so that
the cuts performed, and later on the ranking of candidates from different decay channels,
are meaningful and correct. Had we used the original signal to background ratio during
the trainings, this would have been automatically correct. Because of a too high background level, this was not possible for some channel trainings, so that we artificially
increased the signal component to reach at least 10\%. This resulted in the need to recalculate
the NeuroBayes output after the classification to account for the artificially enhanced
signal component during training as explained in section~\ref{sec:nb_probability}.

\subsection{The Third and Fourth Stage}
\label{chap:3rd_and_4th_stage}
The same procedure of preselection, training and recalculating was then repeated for $D_{(s)}^{*\pm}$ 
and $D^{*0}$ mesons in stage 3 (for the channels and their branching ratios, see table~\ref{table:channels_s3}) 
and finally for $B^{\pm}$ and $B^{0}$ mesons in stage 4.
Variables with good discrimination power were again the product of the NeuroBayes outputs of the children,
the mass of the $D$ meson, the mass difference of the $D$ and $D^{\ast}$ meson and for $B$ meson decays 
the energy-difference $\Delta E$, the angle between the $B$ meson and the thrust axis and angles between
pairs of children.
A list of all used $B$ meson decay modes and the corresponding branching ratios can be seen in 
table~\ref{table:channels_s4}.
The results of the $B^{\pm}$ and $B^{0}$ meson trainings were used to rank the candidates
in each event according to their NeuroBayes outputs. The best candidate selection is now simply
a matter of choosing only the first rank. 

\begin{table}
  \centering
  \begin{tabular}{|llr||llr|}
    \hline
    \multicolumn{3}{|c||}{$D^{\ast +}$} & \multicolumn{3}{c|}{$D^{\ast 0}$} \\ \hline \hline
    \multicolumn{2}{|c}{mode}& \multicolumn{1}{c||}{BR} &\multicolumn{2}{c}{mode} & \multicolumn{1}{c|}{BR} \\ \hline
    $D^{*+} \ \ \rightarrow$ & $D^{0} \pi^{+}$ & $67.70\%$ & $D^{*0} \ \ \rightarrow$ & $D^{0} \pi^{0}$  & $61.90\%$ \\ \hline
    $D^{*+} \ \ \rightarrow$ & $D^{+} \pi^{0}$ & $30.70\%$ & $D^{*0} \ \ \rightarrow$ & $D^{0} \gamma$   & $38.10\%$ \\ \hline
    
    \hline \hline
    
    \multicolumn{3}{|c||}{$D^{\ast}_s$} & & & \\ \hline \hline
    \multicolumn{2}{|c}{mode}& \multicolumn{1}{c||}{BR} &\multicolumn{2}{c}{mode} & \multicolumn{1}{c|}{BR} \\ \hline
    $D_{s}^{+*} \ \ \rightarrow$ & $D_{s}^{+} \gamma$ & $94.20\%$ & & & \\ \hline 

  \end{tabular}
  \caption{Stage 3 - all $D^{*}$ modes (BR from \cite{PDG})}
  \label{table:channels_s3}
\end{table}

\begin{table}
  \centering
  \begin{tabular}{|llr||llr|}
    \hline
    \multicolumn{3}{|c||}{$B^{+}$} & \multicolumn{3}{c|}{$B^{0}$}  \\ \hline \hline
    \multicolumn{2}{|c}{mode}& \multicolumn{1}{c||}{BR} &\multicolumn{2}{c}{mode} & \multicolumn{1}{c|}{BR} \\ \hline
    $B^{+} \ \ \rightarrow$ & $\bar{D}^{0} \pi^{+}$                          &$0.484\%$ &       $B^{0} \ \ \rightarrow$ & $D^{-} \pi^{+}$                       &$0.268\%$ \\ \hline                                
    $B^{+} \ \ \rightarrow$ & $\bar{D}^{0} \pi^{+} \pi^{0}$                  &$1.340\%$&       $B^{0} \ \ \rightarrow$ & $D^{-} \pi^{+} \pi^{0}$                 &$0.760\%$\\ \hline                        
    $B^{+} \ \ \rightarrow$ & $\bar{D}^{0} \pi^{+} \pi^{+} \pi^{-}$          &$1.100\%$&       $B^{0} \ \ \rightarrow$ & $D^{-} \pi^{+} \pi^{+} \pi^{-}$          &$0.800\%$\\ \hline                
    $B^{+} \ \ \rightarrow$ & $D_{s}^{+} \bar{D}^{0}$                        &$1.000\%$&       $B^{0} \ \ \rightarrow$ & $\bar{D}^{0} \pi^{0}$                   &$0.026\%$\\ \hline                          
    $B^{+} \ \ \rightarrow$ & $\bar{D}^{0*} \pi^{+}$                         &$0.519\%$&       $B^{0} \ \ \rightarrow$ & $D_{s}^{+} D^{-}$                       &$0.720\%$\\ \hline                              
    $B^{+} \ \ \rightarrow$ & $\bar{D}^{0*} \pi^{+} \pi^{0}$                 &$0.980\%$&       $B^{0} \ \ \rightarrow$ & $D^{*-} \pi^{+}$                        &$0.276\%$\\ \hline                               
    $B^{+} \ \ \rightarrow$ & $\bar{D}^{0*} \pi^{+} \pi^{+} \pi^{-}$         &$1.030\%$&       $B^{0} \ \ \rightarrow$ & $D^{*-} \pi^{+} \pi^{0}$                 &$1.500\%$\\ \hline                       
    $B^{+} \ \ \rightarrow$ & $\bar{D}^{0*} \pi^{+} \pi^{+} \pi^{-} \pi^{0}$ &$1.800\%$&       $B^{0} \ \ \rightarrow$ & $D^{*-} \pi^{+} \pi^{+} \pi^{-}$          &$0.700\%$\\ \hline               
    $B^{+} \ \ \rightarrow$ & $D_{s}^{+*} \bar{D}^{0}$                       &$0.760\%$&       $B^{0} \ \ \rightarrow$  & $D^{*-} \pi^{+} \pi^{+} \pi^{-} \pi^{0}$ &$1.760\%$\\ \hline      
    $B^{+} \ \ \rightarrow$ & $D_{s}^{+} \bar{D}^{0*}$                       &$0.820\%$&       $B^{0} \ \ \rightarrow$ & $D_{s}^{+*} D^{-}$                       &$0.740\%$\\ \hline                             
    $B^{+} \ \ \rightarrow$ & $D_{s}^{+*} \bar{D}^{0*}$                      &$1.710\%$&       $B^{0} \ \ \rightarrow$ & $D_{s}^{+} D^{*-}$                       &$0.800\%$\\ \hline                             
    $B^{+} \ \ \rightarrow$ & $\bar{D}^{0} K^{+}$                            &$0.037\%$&      $B^{0} \ \ \rightarrow$ & $D_{s}^{+*} D^{*-}$                      &$1.770\%$\\ \hline                            
    $B^{+} \ \ \rightarrow$ & $D^{-} \pi^{+} \pi^{+}$                        &$0.107\%$&       $B^{0} \ \ \rightarrow$ & $J/\psi K_{S}^{0}$                       &$0.087\%$\\ \hline                             
    $B^{+} \ \ \rightarrow$ & $J/\psi K^{+}$                                 &$0.101\%$&       $B^{0} \ \ \rightarrow$ & $J/\psi K^{+} \pi^{-}$                   &$0.120\%$\\ \hline                         
    $B^{+} \ \ \rightarrow$ & $J/\psi K^{+} \pi^{+} \pi^{-}$                 &$0.107\%$&       $B^{0} \ \ \rightarrow$ & $J/\psi K_{S}^{0} \pi^{+} \pi^{-}$        &$0.100\%$\\ \hline             
    $B^{+} \ \ \rightarrow$ & $J/\psi K^{+} \pi^{0}$                         &$0.047\%$& & &\\ \hline 
    $B^{+} \ \ \rightarrow$ & $J/\psi K_{S}^{0} \pi^{+}$                     &$0.094\%$& & &\\ \hline 
  \end{tabular}
  \caption{Stage 4 - All $B$ modes (BR from \cite{PDG})}
  \label{table:channels_s4}
\end{table}

\subsection{Suppression of non $B\bar{B}$ Background}
\label{chap:continuum_suppression}
Non $B\bar{B}$ events differ from $B\bar{B}$ events in the event shape. As there is hardly any kinetic energy in $B\bar{B}$ events left, the decay particles are much more spherically distributed in contrast to the jet-like structure of non-$B\bar{B}$ events.
There are numerous variables to quantify the different event shapes. The reduced second Fox-Wolfram Moment 
$R_2$~\cite{PhysRevLett.41.1581} gives non-candidate-specific information about the event shape, the thrust 
angle and $\cos \Theta_B$ provide
information for each individual candidate. The Super Fox-Wolfram 
Moments (SFWM)~\cite{PhysRevLett.91.261801} contain additional information about 
the tag- and signal-side. 

In the default mode of the full reconstruction, no event shape variables are used, 
as this might introduce some bias for certain analyses.
There is, however, an additional algorithm that can be used after the full reconstruction. 
This algorithm recalculates the NeuroBayes output of all of the $B$ candidates and determines the best candidate 
again, based on the new output.
The algorithm incorporates two continuum suppression networks. The first network uses the reduced second 
Fox-Wolfram Moment, the thrust angle and $\cos \Theta_B$. It therefore only depends on $B_{\rm tag}$.
The second network additionally contains the Super Fox-Wolfram Moments, depending also on $B_{\rm sig}$.
As these networks take more information into account, there is a significant improvement in the
quality of the NeuroBayes output and also in the best candidate selection. The results 
can be found in figures~\ref{fig:pur_eff_bp} and~\ref{fig:pur_eff_b0}.


\section{Performance of the new Algorithm}
\subsection{Efficiency Estimation}
There is an existing full reconstruction algorithm at Belle 
(see e.g. \cite{PhysRevD.77.091503,PhysRevLett.99.221802,PhysRevLett.97.251802,PhysRevD.72.051109,PhysRevLett.95.241801}), using a classical, cut-based reconstruction method without taking probabilistic information into account.
We compare the performance of the new and the classical algorithm  
by estimating the numbers of correctly reconstructed $B_{\rm tag}$ candidates using the final Belle data sample 
collected at the $\Upsilon(4S)$ resonance. 

The sample contains $771.6 \times 10^{6}$ $B\bar{B}$ pairs. The kinematic consistency of a $B_{\rm tag}$ 
candidate with a $B$ meson decay is checked using the beam-energy constrained mass 
$M_{bc}\equiv \sqrt{E^2_{\rm beam} - p^2_{B}} $, where $E_{\rm beam}$ is the measured beam energy and $p_{B}$ 
is the reconstructed four-momentum of the $B$ meson in the center-of-mass rest frame. 
None of the variables used in the network trainings are correlated with $M_{bc}$, 
which can therefore be used to estimate the number of correctly reconstructed $B_{\rm tag}$ candidates from
fits to the $M_{bc}$ distribution. 

Since this paper focuses on the description of the new full reconstruction method and its improvements, 
we do not evaluate systematic uncertainties on the fitted signal yields as would be required for 
physics analyses.
Typically any full reconstruction tool is used in conjunction with a signal side analysis. For most signal side
analyses, only the largest possible efficiency of the tag side sample is important, as the background is
reduced dramatically by the signal side selection. When we want to compare two full reconstruction methods by themselves, 
without any signal side selection, we have to perform a fit to the inclusive tag side  $M_{bc}$ distribution.
Especially for the new full reconstruction, this distribution contains large amounts of background, which are 
irrelevant for most analyses, but make the fit results less reliable. Therefore it is only possible to give a 
quite raw estimate for the signal gain and therefore for the improvement compared to the classical full reconstruction
for maximum efficiency.
This is usually not a problem for physics analyses because of the applied selection criteria.

The number of correctly reconstructed $B$ mesons are estimated from the fit to be 2.1 million $B^{\pm}$ 
and 1.4 million $B^{0}$ for the maximum efficiency case.
This corresponds to an efficiency of roughly $0.18\%$ for $B^{0}$ 
and $0.28\%$ for $B^{\pm}$. This efficiency is defined as the number of correct reconstructed $B$ mesons divided 
by the number of produced $B\bar{B}$ pairs, which is the same as the number of produced $B^{0}$ and $B^{\pm}$ 
mesons respectively. Note that in other publications a different definition might be used, which takes the 
number of produced charged or neutral $B$ meson pairs as normalization, resulting in twice the value for the 
single $B$ meson reconstruction efficiency.

\begin{figure}[htbp]
  \centering
  \includegraphics[width=\linewidth]{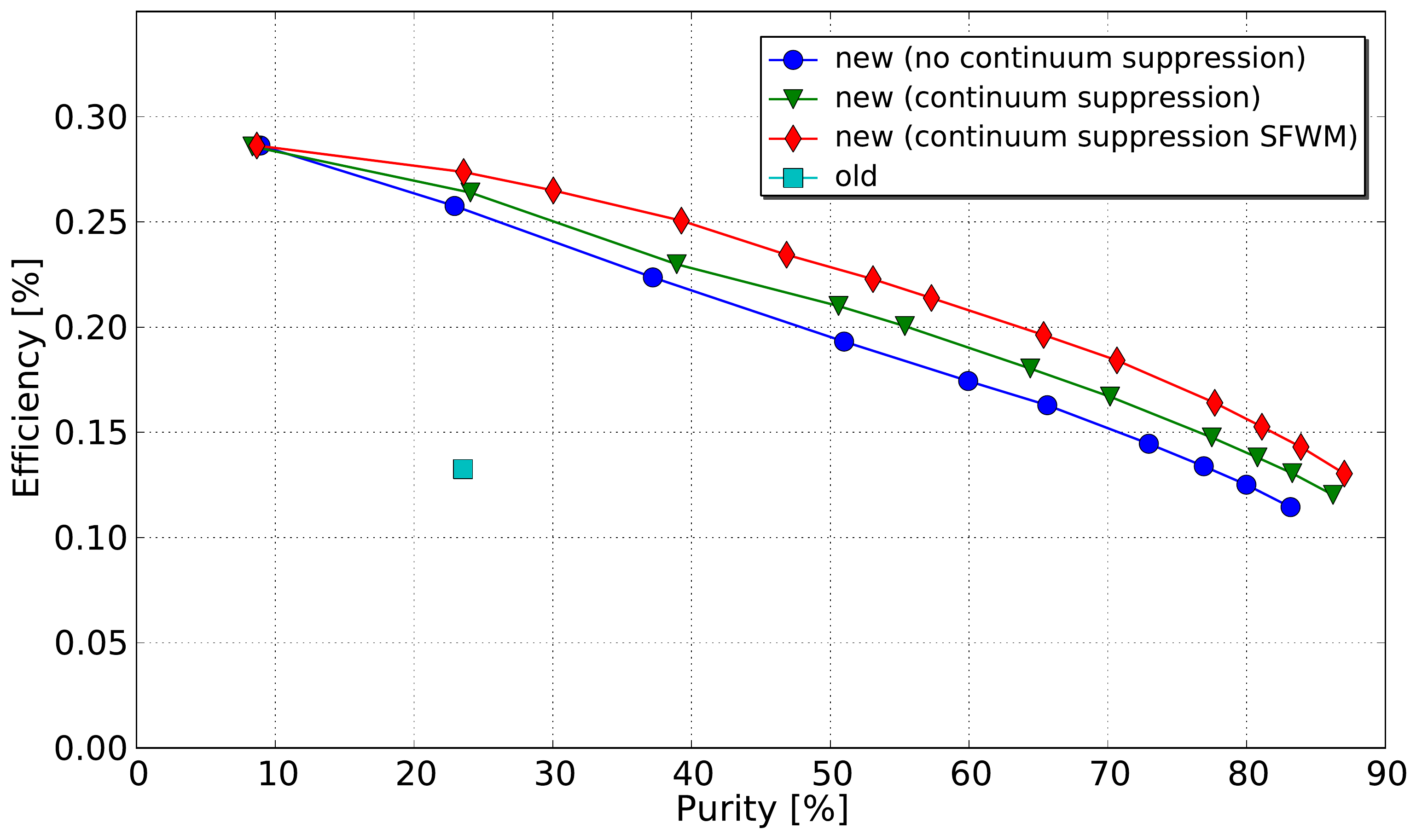}
  \caption{Purity-efficiency plot for $B^{+}$ mesons}
  \label{fig:pur_eff_bp}
\end{figure}

\begin{figure}[htbp]
  \centering
  \includegraphics[width=\linewidth]{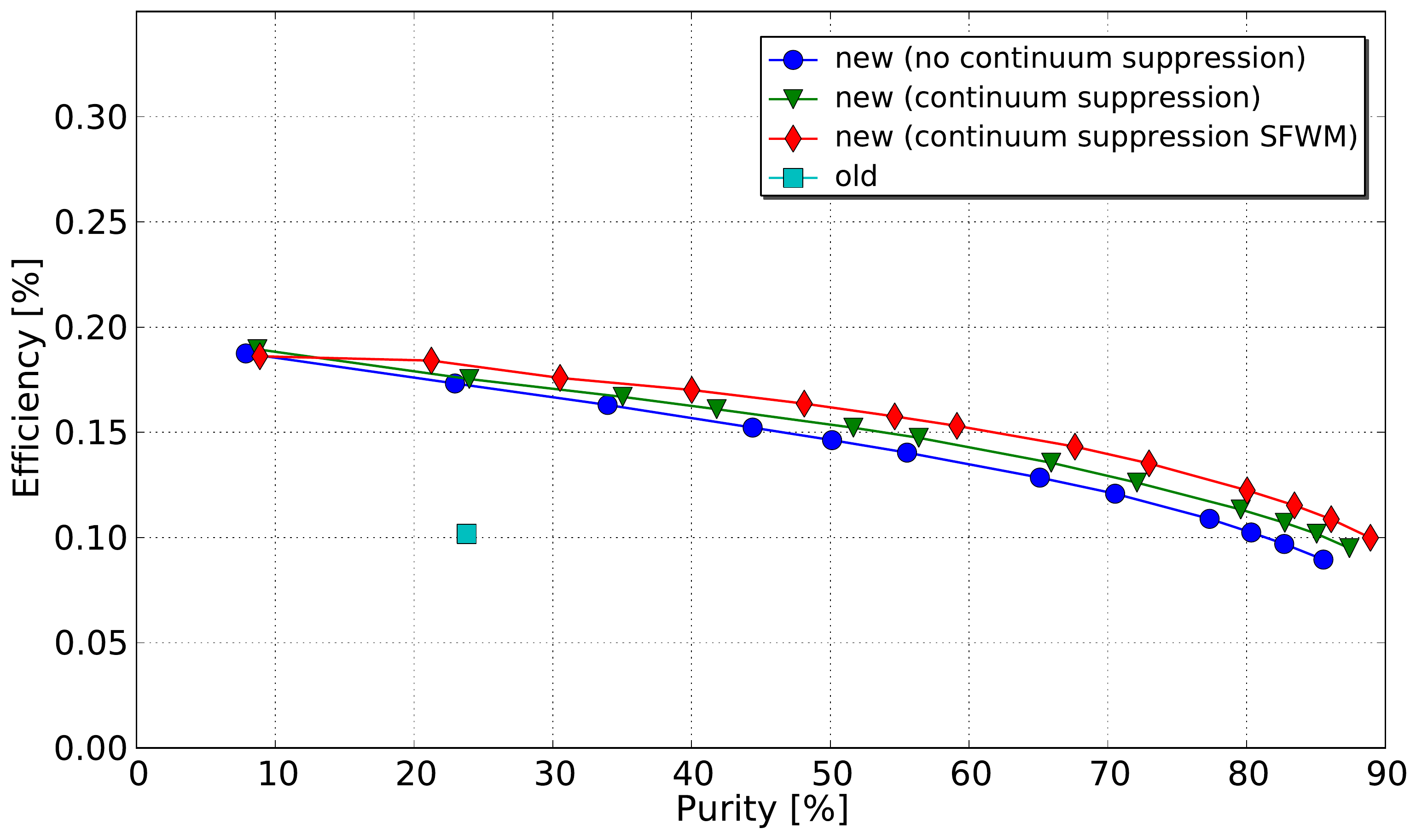}
  \caption{Purity-efficiency plot for $B^{0}$ mesons}
  \label{fig:pur_eff_b0}
\end{figure}

In order to get more reliable fit results, we can introduce cuts on the NeuroBayes outputs of the $B^{\pm}$ 
and $B^{0}$ meson networks, and thereby choose efficiency and purity freely. 
Figures~\ref{fig:pur_eff_bp} and~\ref{fig:pur_eff_b0} show the resulting purity-efficiency plots for the three modes explained in chapter~\ref{chap:continuum_suppression}. Purity is defined as the ratio of the signal 
component of the fit to the entire fit result integrated over the region $M_{bc} > 5.27 $GeV$/c^2$.

\begin{figure}[htbp]
  \centering
  \subfigure[$B^+$ selection with roughly equal purity]
  {
   \includegraphics[width=0.45\linewidth]{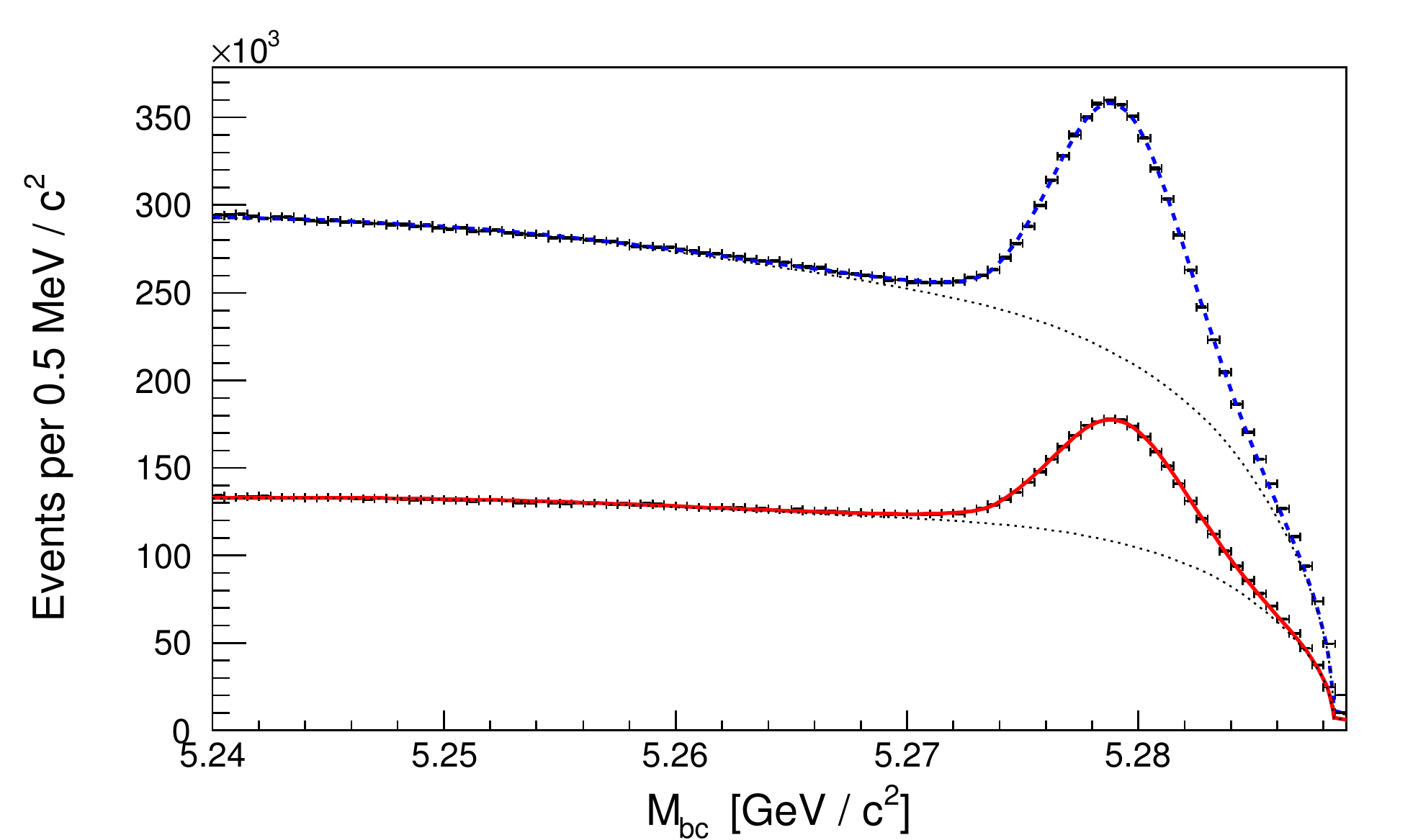}
   \label{fig:bp_same_purity}
  }
  \subfigure[$B^{+}$ selection with roughly equal background level]
  {
   \includegraphics[width=0.45\linewidth]{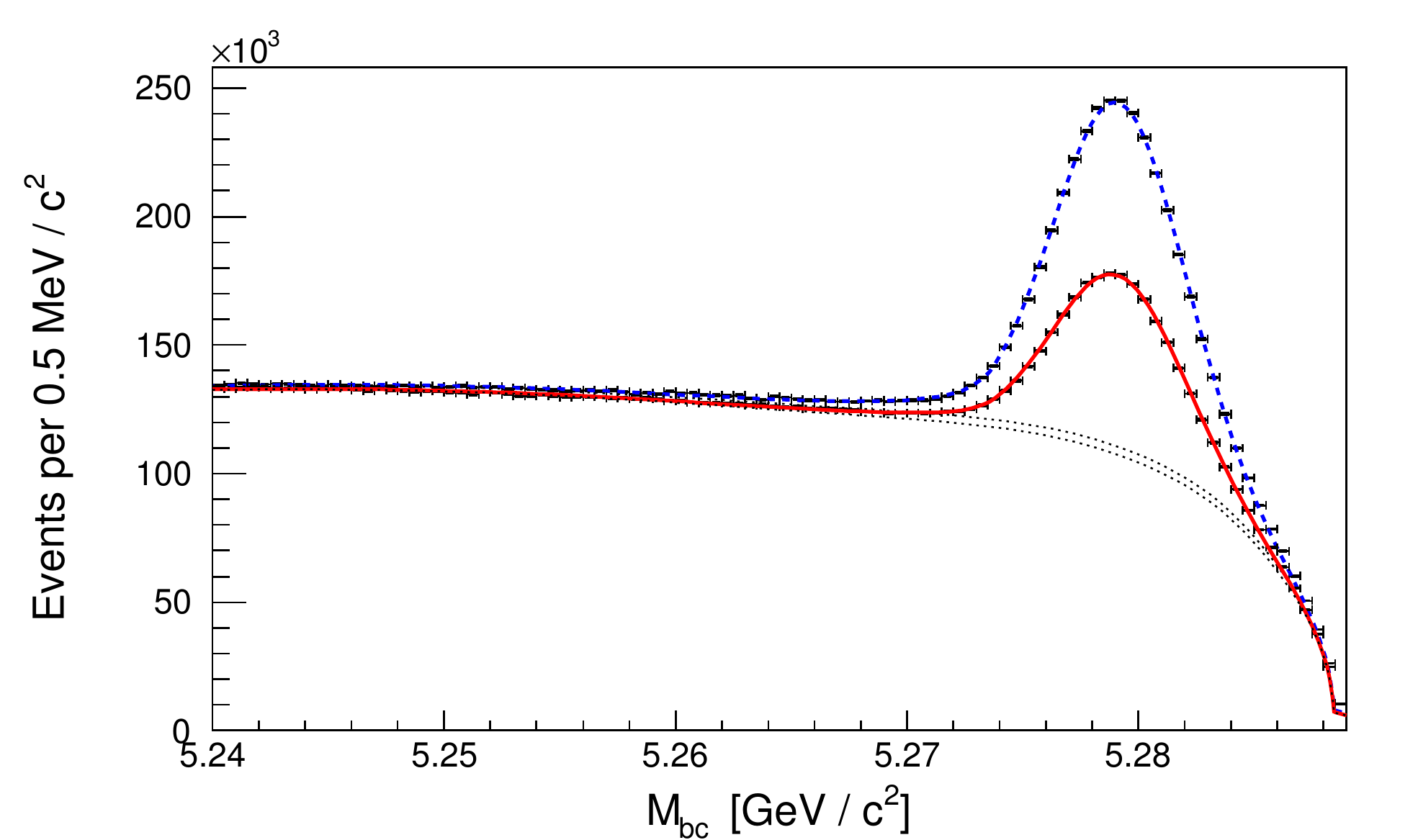}
   \label{fig:bp_same_background}
  }
  \subfigure[$B^+$ selection with roughly equal efficiency]
  {
   \includegraphics[width=0.45\linewidth]{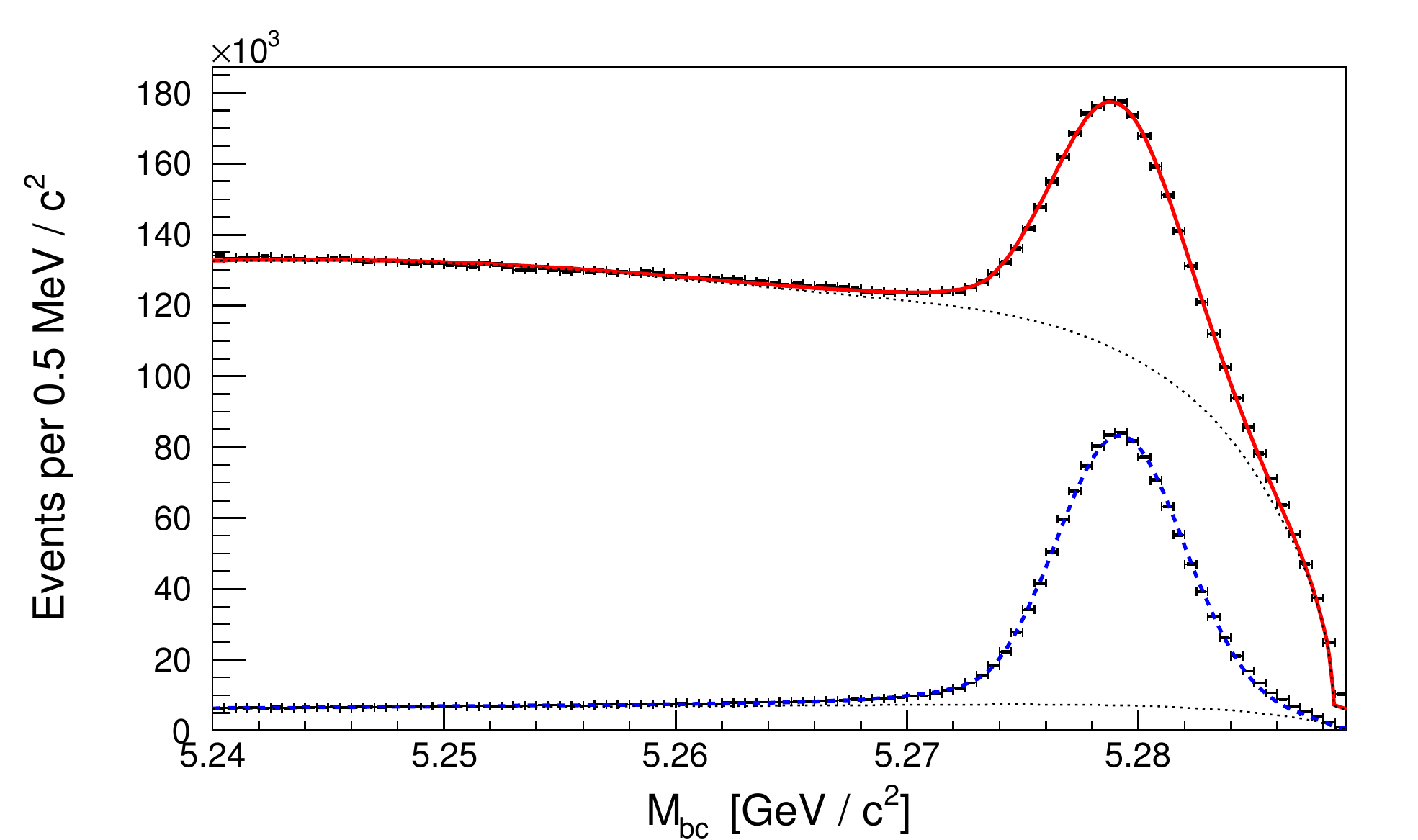}
   \label{fig:bp_same_efficiency}
  }
  \subfigure[$B^{0}$ selection with roughly equal efficiency]
  {
   \includegraphics[width=0.45\linewidth]{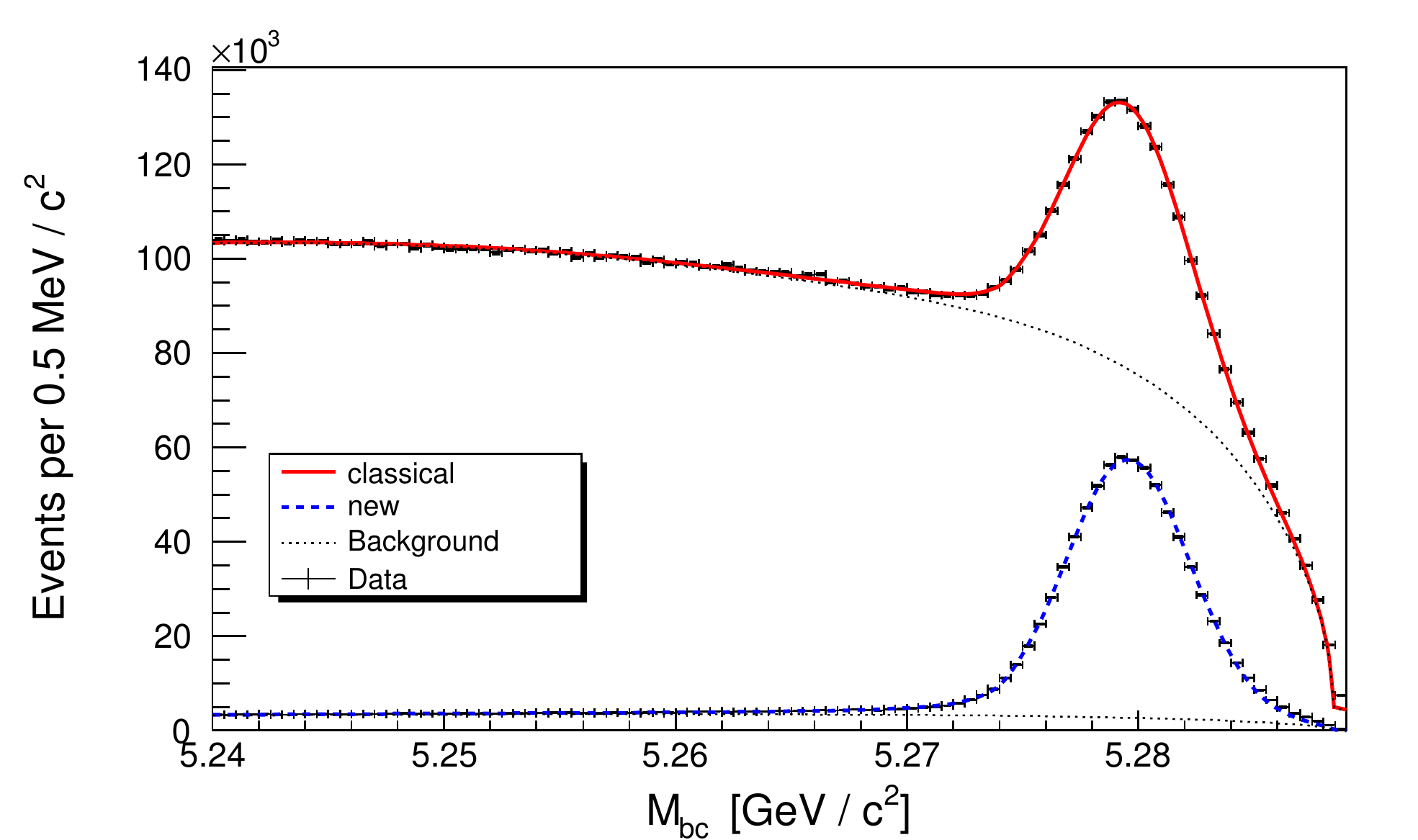}
   \label{fig:b0_same_efficiency}
  }
  \caption[Optional caption for list of figures]{$M_{bc}$ plots for different selections: The dashed blue line is a fit of the $M_{bc}$ distributions for the new full reconstruction algorithm, the solid red line to the classical one. The network cuts are chosen to have \subref{fig:bp_same_purity} roughly equal purity, \subref{fig:bp_same_background} roughly equal background level, \subref{fig:bp_same_efficiency}, \subref{fig:b0_same_efficiency} roughly equal efficiency compared to the classical one}
  \label{fig:mbc_plots}
\end{figure}

If no cut is performed, the standard selection that gives maximum efficiency is used. One can also choose a cut, corresponding to the same purity as in the classical full reconstruction tool, which results
in an increase of efficiency by approximately a factor of $2$ , as shown in figure~\ref{fig:bp_same_purity}. 
A cut, corresponding to the same background level is shown in figure~\ref{fig:bp_same_background}.  
One is also free to choose the
same efficiency as in the classical full reconstruction. This results in an increase in the purity from
about $25\%$ to nearly $90\%$ as shown in figures~\ref{fig:bp_same_efficiency} and~\ref{fig:b0_same_efficiency}. 
Any working point between and even beyond
these three examples can be chosen in a very simple manner (cutting on the output of of the stage 4 networks) by the user.



\subsection{Without new Channels}
\label{chap:without_new_channels}
If we exclude the newly added $D$ and $B$ decay channels from the full reconstruction and choose 
a network output cut to achieve the same background level as in the classical full
reconstruction, the efficiency is increased by approximately 50\% for $B^{0}$ mesons
and 60\% for $B^{+}$ mesons. A comparison of the individual $B$ decay channels revealed that 
the largest improvement was achieved in modes with two or more light mesons, where the
new full reconstruction does not impose any phase-space limits.
The newly added channels make a valuable contribution of approximately $20\%$ 
of the entire signal sample for both $B^{0}$ and $B^{+}$ mesons.

\subsection{Applied Example: Missing Mass Reconstruction}
\label{sec:dstarlnu}
\begin{figure}[htbp]
  \centering
  \subfigure[\ $B^{0} \to D^{\ast -} \ \ell^{+} \ \nu_{\ell}$ - new full reconstruction]{
    \includegraphics[width=0.40\linewidth, angle=0]{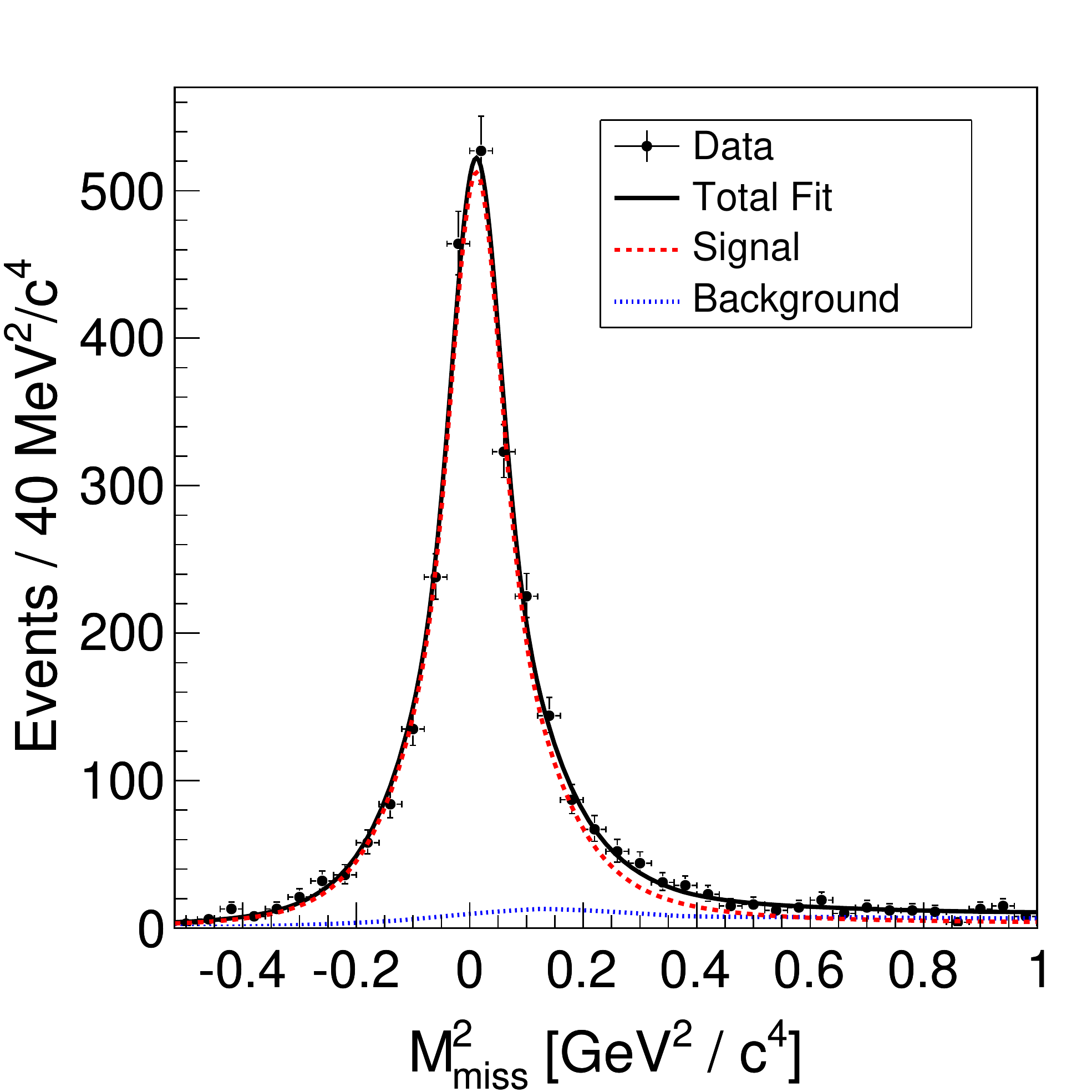}
    \label{fig:b0_dstlnu_mm2_new}
  }
  \subfigure[\ $B^{0} \to D^{\ast -} \ \ell^{+} \ \nu_{\ell}$ - classical full reconstruction]{
    \includegraphics[width=0.40\linewidth, angle=0]{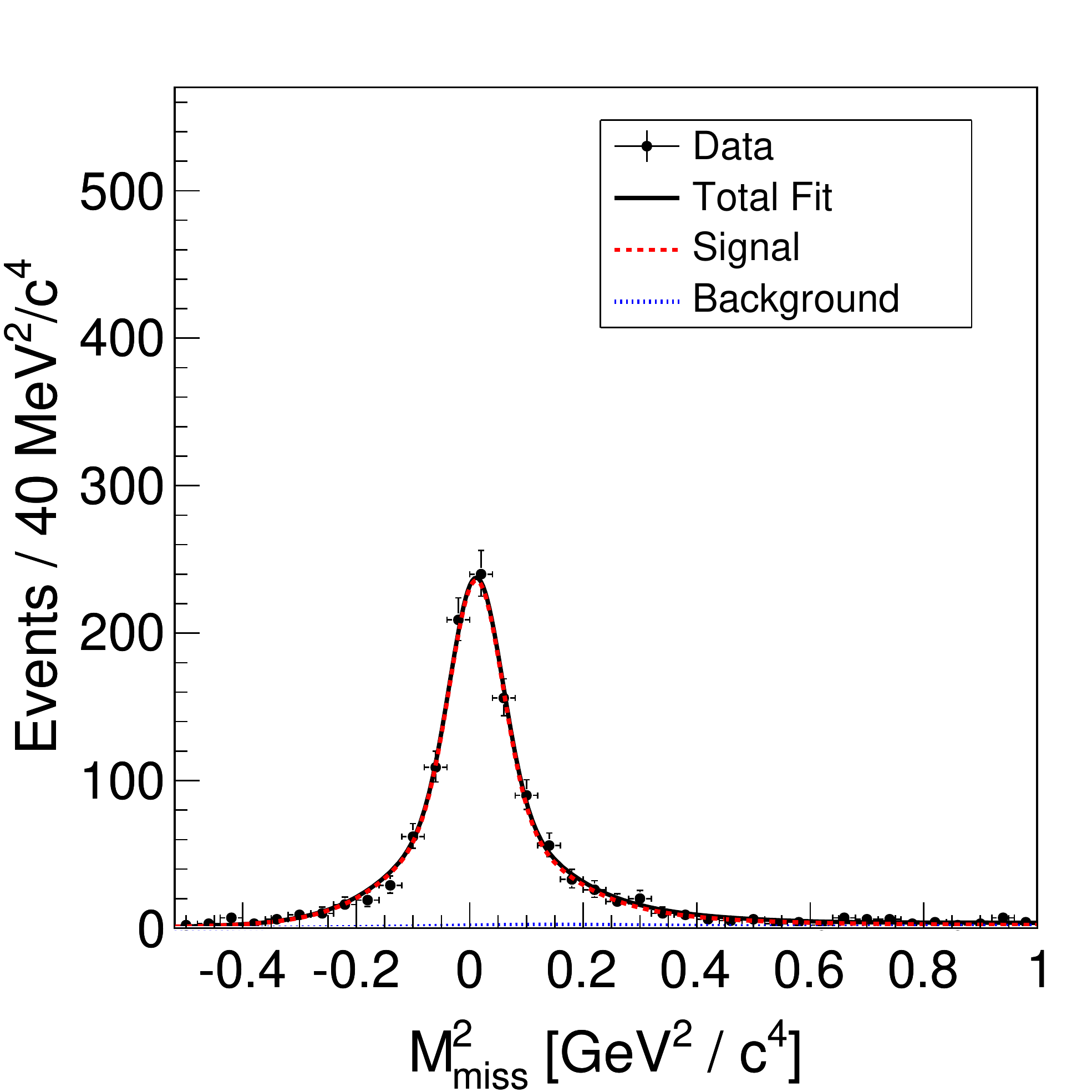}
    \label{fig:b0_dstlnu_mm2_old}
  }
  \caption[Optional caption for list of figures]{Missing mass distributions for $B^{0} \to D^{\ast -} \ \ell^{+} \ \nu_{\ell}$ decays of the new and classical full reconstruction tool}
  \label{fig:mm2_b0}
\end{figure}

In order to test the results of the full reconstruction and also
to compare the performance to its predecessor, a quick benchmark
analysis was performed. This was the search for the decay

\begin{align}
  B^{0} &\rightarrow D^{\ast -} \ell^{+} \nu_{\ell} 
\label{eqn:dstarlnu_four_signals}
\end{align}
on the signal side. A kinematic variable used to distinguish correctly reconstructed signal candidates from background candidates is the missing mass, defined as

\begin{align}
 M^2_{\rm miss}=\lvert p_{\Upsilon(4S)} - ( \sum_{i} p_{i} + p_{B_{\rm tag}} ) \rvert^2 , 
\end{align}
where $p_{\Upsilon(4S)}$ denotes the four-momentum of the $\Upsilon(4s)$ resonance, $p_{B_{\rm tag}}$ is the four-momentum 
of the $B_{\rm tag}$ and $\sum_{i} p_{i}$ is the sum of the four-momenta of the reconstructed particles on the signal side.
Because the neutrino is the only missing particle in this decay, we expect the missing mass to be zero for signal events. 
The result can be seen in figure~\ref{fig:b0_dstlnu_mm2_new} for the new full reconstruction algorithm and as an 
comparison in  figure~\ref{fig:b0_dstlnu_mm2_old} the result for the classical full reconstruction algorithm. 
A clear peak is observed at the expected position with similar resolutions for new and classical full reconstruction.
Thus despite the addition of less clean decay modes, the momentum resolution of the fully reconstructed $B$ meson is preserved. As expected we also observe in this applied example a significant increase of efficiency.

\section{Conclusion}
\label{sec:conclusion}
We have developed an improved full reconstruction algorithm for the Belle experiment by introducing a hierarchical selection procedure. Instead of cutting away candidates in the early stages, we postpone the decision to later stages by very soft selections on the product of their Bayesian signal probability and giving this probability as an input for the higher stages networks. Together with a higher separation power of the neural networks compared to a cut based selection, this enabled us to reconstruct more decay channels with an acceptable computation time. Depending on the analysis, we expect an overall improvement of the effective luminosity of roughly a factor of 2 for a large number of analyses relying on the full reconstruction.

\section{Acknowledgments}
We thank Toru Iijima, Karim Trabelsi and Gary Barker for their careful reading of this article.


\bibliographystyle{model1a-num-names}
\bibliography{bibliography_db}

\begin{thebibliography}{14}
\expandafter\ifx\csname natexlab\endcsname\relax\def\natexlab#1{#1}\fi
\providecommand{\bibinfo}[2]{#2}
\ifx\xfnm\relax \def\xfnm[#1]{\unskip,\space#1}\fi
\bibitem[{Kurokawa and Kikutani(2003)}]{KEKB}
\bibinfo{author}{S.~Kurokawa}, \bibinfo{author}{E.~Kikutani},
\newblock \bibinfo{title}{{Overview of the KEKB accelerators}},
\newblock \bibinfo{journal}{Nuclear Instruments and Methods in Physics Research
  Section A: Accelerators, Spectrometers, Detectors and Associated Equipment}
  \bibinfo{volume}{499} (\bibinfo{year}{2003}) \bibinfo{pages}{1--7}.
\bibitem[{Abashian et~al.(2002)}]{Belle_detector}
\bibinfo{author}{A.~Abashian}, et~al.,
\newblock \bibinfo{title}{{The Belle detector}},
\newblock \bibinfo{journal}{Nuclear Instruments and Methods in Physics Research
  Section A: Accelerators, Spectrometers, Detectors and Associated Equipment}
  \bibinfo{volume}{479} (\bibinfo{year}{2002}) \bibinfo{pages}{117--232}.
\bibitem[{Nakamura and {Particle Data Group}(2010)}]{PDG}
\bibinfo{author}{K.~Nakamura}, \bibinfo{author}{{Particle Data Group}},
\newblock \bibinfo{title}{{Review of Particle Physics}},
\newblock \bibinfo{journal}{Journal of Physics G: Nuclear and Particle Physics}
  \bibinfo{volume}{37} (\bibinfo{year}{2010}) \bibinfo{pages}{075021}.
\bibitem[{Feindt(2004)}]{neurobayes}
\bibinfo{author}{M.~Feindt},
\newblock \bibinfo{title}{{A Neural Bayesian Estimator for Conditional
  Probability Densities}},
\newblock \bibinfo{journal}{{arXiv:physics/0402093}}  (\bibinfo{year}{2004}).
\bibitem[{Allison et~al.(2006)}]{GEANT}
\bibinfo{author}{J.~Allison}, et~al.,
\newblock \bibinfo{title}{{Geant4 developments and applications}},
\newblock \bibinfo{journal}{Nuclear Science, IEEE Transactions on}
  \bibinfo{volume}{53} (\bibinfo{year}{2006}) \bibinfo{pages}{270 --278}.
\bibitem[{Sj{\"o}strand et~al.(2006)Sj{\"o}strand, Mrenna, and Skands}]{PYTHIA}
\bibinfo{author}{T.~Sj{\"o}strand}, \bibinfo{author}{S.~Mrenna},
  \bibinfo{author}{P.~Skands},
\newblock \bibinfo{title}{Pythia 6.4 physics and manual},
\newblock \bibinfo{journal}{Journal of High Energy Physics}
  \bibinfo{volume}{2006} (\bibinfo{year}{2006}) \bibinfo{pages}{026}.
\bibitem[{Lange(2001)}]{Lange:2001}
\bibinfo{author}{D.~Lange},
\newblock \bibinfo{title}{{The EvtGen particle decay simulation package}},
\newblock \bibinfo{journal}{Nucl.~Instrum.~Meth.} \bibinfo{volume}{A462}
  (\bibinfo{year}{2001}) \bibinfo{pages}{152--155}.
\bibitem[{Fox and Wolfram(1978)}]{PhysRevLett.41.1581}
\bibinfo{author}{G.~C. Fox}, \bibinfo{author}{S.~Wolfram},
\newblock \bibinfo{title}{{Observables for the Analysis of Event Shapes in
  $e^{+}e^{-}$ Annihilation and Other Processes}},
\newblock \bibinfo{journal}{Phys. Rev. Lett.} \bibinfo{volume}{41}
  (\bibinfo{year}{1978}) \bibinfo{pages}{1581--1585}.
\bibitem[{Lee et~al.(2003)}]{PhysRevLett.91.261801}
\bibinfo{author}{S.~H. Lee}, et~al.,
\newblock \bibinfo{title}{{Evidence for
  $B^{0}\rightarrow{}\pi{}^{0}\pi{}^{0}$}},
\newblock \bibinfo{journal}{Phys. Rev. Lett.} \bibinfo{volume}{91}
  (\bibinfo{year}{2003}) \bibinfo{pages}{261801}.
\bibitem[{Liventsev et~al.(2008)}]{PhysRevD.77.091503}
\bibinfo{author}{D.~Liventsev}, et~al.,
\newblock \bibinfo{title}{{Study of $B\rightarrow{}D^{**}l\nu{}$ with full
  reconstruction tagging}},
\newblock \bibinfo{journal}{Phys. Rev. D} \bibinfo{volume}{77}
  (\bibinfo{year}{2008}) \bibinfo{pages}{091503}.
\bibitem[{Chen et~al.(2007)}]{PhysRevLett.99.221802}
\bibinfo{author}{K.-F. Chen}, et~al.,
\newblock \bibinfo{title}{{Search for $B\rightarrow{}h^{(*)}\nu{}\nu{}$ Decays
  at Belle}},
\newblock \bibinfo{journal}{Phys. Rev. Lett.} \bibinfo{volume}{99}
  (\bibinfo{year}{2007}) \bibinfo{pages}{221802}.
\bibitem[{Ikado et~al.(2006)}]{PhysRevLett.97.251802}
\bibinfo{author}{K.~Ikado}, et~al.,
\newblock \bibinfo{title}{{Evidence of the Purely Leptonic Decay
  $B^{-}\rightarrow{}\tau{}^{-}\nu{}_{\tau{}}$}},
\newblock \bibinfo{journal}{Phys. Rev. Lett.} \bibinfo{volume}{97}
  (\bibinfo{year}{2006}) \bibinfo{pages}{251802}.
\bibitem[{Liventsev et~al.(2005)}]{PhysRevD.72.051109}
\bibinfo{author}{D.~Liventsev}, et~al.,
\newblock \bibinfo{title}{{Measurement of the branching fractions for
  $B^{-}\rightarrow{}D^{(*)+}\pi{}^{-}\ell{}^{-}\nu{}_{\ell{}}$ and
  $B^{0}\rightarrow{}D^{(*)0}\pi{}^{+}\ell{}^{-}\nu{}_{\ell{}}$}},
\newblock \bibinfo{journal}{Phys. Rev. D} \bibinfo{volume}{72}
  (\bibinfo{year}{2005}) \bibinfo{pages}{051109}.
\bibitem[{Bizjak et~al.(2005)}]{PhysRevLett.95.241801}
\bibinfo{author}{I.~Bizjak}, et~al.,
\newblock \bibinfo{title}{{Determination of $|V_{ub}|$ from Measurements of the
  Inclusive Charmless Semileptonic Partial Rates of $B$ Mesons using Full
  Reconstruction Tags}},
\newblock \bibinfo{journal}{Phys. Rev. Lett.} \bibinfo{volume}{95}
  (\bibinfo{year}{2005}) \bibinfo{pages}{241801}.

\end{thebibliography}

\end{document}